\documentclass[11pt]{article}
\usepackage{axodraw}
\usepackage{epsfig}
\usepackage{amsfonts}
\usepackage{amsmath}
\usepackage{bm,bbm}
\usepackage{cite}
\allowdisplaybreaks[1]

\input pix.sty

\def\wv#1{\widetilde{#1}}
\newcommand{\A}[1]{A^{ }_\rmii{$#1$}}
\newcommand{\Ad}[1]{\dot A^{ }_\rmii{$#1$}}
\renewcommand{\B}[2]{B^{ }_\rmii{$#1#2$}}
\newcommand{\Bd}[2]{\dot B^{ }_\rmii{$#1#2$}}
\newcommand{\BT}[2]{B^{ }_\rmii{$T,\!#1#2$}}
\newcommand{\D}{\Delta}

\newcommand{\diag}{\mathop{\mbox{diag}}}

\newcommand{\E}{\rmii{$E$}}
\newcommand{\M}{\rmii{$M$}}

\newcommand{\sL}{\rmii{$L$}}

\newcommand{\T}{\rmii{$T$}}
\newcommand{\W}{\rmii{$W$}}

\newcommand{\Z}{\rmii{$Z$}}
\newcommand{\Wt}{\rmii{$\widetilde W$}}
\newcommand{\Zt}{\rmii{$\widetilde Z$}}
\newcommand{\Qt}{\rmii{$\widetilde Q$}}

\newcommand{\mW}{m_\rmii{$W$}}
\newcommand{\mZ}{m_\rmii{$Z$}}
\newcommand{\mQ}{m_\rmii{$Q$}}
\newcommand{\mWt}{m_\rmii{$\widetilde W$}}
\newcommand{\mZt}{m_\rmii{$\widetilde Z$}}
\newcommand{\mQt}{m_\rmii{$\widetilde Q$}}
\newcommand{\mH}{m_\rmii{$H$}}

\newcommand{\aL}{a^{ }_\rmii{L}}

\renewcommand{\eq}{eq.~}
\renewcommand{\eqs}{eqs.~}
\renewcommand{\se}{sec.~}

\renewcommand{\fig}{fig.~}

\newcommand{\rmO}{{\mathcal{O}}}

\def\lsi{\raise0.3ex\hbox{$<$\kern-0.75em\raise-1.1ex\hbox{$\sim$}}}
\def\gsi{\raise0.3ex\hbox{$>$\kern-0.75em\raise-1.1ex\hbox{$\sim$}}}

\newcommand{\rmii}[1]{{\mbox{\tiny\rm{#1}}}}

\newcommand{\re}{\mathop{\mbox{Re}}}
\newcommand{\im}{\mathop{\mbox{Im}}}

\newcommand{\Tint}[1]{{\hbox{$\sum$}\!\!\!\!\!\!\!\int\,}_{\!\!\!\!\raise-0.9ex\hbox{$\scriptstyle{#1}$}}}
\newcommand{\Tinti}[1]{{{\Sigma}\!\!\!\!\raise0.3ex\hbox{$\int$}_\rmii{${#1}$}}}

\newcommand{\bi}{\begin{itemize}}
\newcommand{\ei}{\end{itemize}}
\newcommand{\hide}[1]{ }
\newcommand{\bsl}[1]{\,\slash\!\!\!\!{#1}\,}
\newcommand{\msl}[1]{\,\slash\!\!\!{#1}\,}
\def\TAsc(#1,#2)(#3,#4,#5)%
{\SetWidth{2.0}\CArc(#1,#2)(#3,#4,#5)\SetWidth{1.0}}
\def\Lwidth{3}

\def\TAgl(#1,#2)(#3,#4,#5){\SetWidth{2.0}\PhotonArc(#1,#2)(#3,#4,#5){\Lwidth}%
{6.283 #3 mul 360 div #4 #5 sub #4 #5 sub mul sqrt mul Tdensity mul}%
\SetWidth{1.0}}
\def\TLgl(#1,#2)(#3,#4){\SetWidth{2.0}\Photon(#1,#2)(#3,#4){\Lwidth}
{#1 #3 sub #1 #3 sub mul #2 #4 sub #2 #4 sub mul add sqrt Tdensity mul}%
\SetWidth{1.0}}

\renewcommand{\picc}[1]{\;\parbox[c]{60pt}{\begin{picture}(60,30)(0,0)
\SetWidth{1.0}\SetScale{1.3} #1 \end{picture}}\; }
\def\Lwidth{1.3}

\renewcommand{\pic}[1]{\;\parbox[c]{30pt}{\begin{picture}(30,30)(0,-3)
\SetWidth{1.0}\SetScale{0.8} #1 \end{picture}}\;}
\renewcommand{\picb}[1]{\;\parbox[c]{45pt}{\begin{picture}(45,30)(0,-3)
\SetWidth{1.0}\SetScale{0.8} #1 \end{picture}}\;}
\def\xprocA{\pic{%
 \Agl(25,15)(15,5,175)%
 \Line(0,14)(50,14)%
 \Line(0,16)(50,16)%
}}
\def\xprocB{\pic{%
 \CArc(25,15)(15,5,175)%
 \Line(0,14)(50,14)%
 \Line(0,16)(50,16)%
}}
\def\xprocC{\pic{%
 \Agl(25,15)(15,5,175)%
 \Line(0,14)(50,14)%
 \Line(0,16)(50,16)%
 \CCirc(25,15){3}{Black}{Black}
}}
\def\xprocD{\pic{%
 \CArc(25,15)(15,5,175)%
 \Line(0,14)(50,14)%
 \Line(0,16)(50,16)%
 \CCirc(25,15){3}{Black}{Black}
}}
\def\yprocA{\pic{%
 \Agl(25,15)(15,5,175)%
 \Line(0,14)(50,14)%
 \Line(0,16)(50,16)%
 \CCirc(25,30){5}{Black}{Gray}
}}
\def\yprocB{\pic{%
 \CArc(25,15)(15,5,175)%
 \Line(0,14)(50,14)%
 \Line(0,16)(50,16)%
 \CCirc(25,30){5}{Black}{Gray}
}}
\def\yprocC{\pic{%
 \Agl(25,15)(15,5,175)%
 \Line(0,14)(50,14)%
 \Line(0,16)(50,16)%
 \CCirc(25,15){3}{Black}{Black}
 \CCirc(25,30){5}{Black}{Gray}
}}
\def\yprocD{\pic{%
 \CArc(25,15)(15,5,175)%
 \Line(0,14)(50,14)%
 \Line(0,16)(50,16)%
 \CCirc(25,15){3}{Black}{Black}
 \CCirc(25,30){5}{Black}{Gray}
}}
\def\yprocEa{\pic{%
 \Agl(25,15)(15,5,175)%
 \Lgl(25,16)(25,29.5)%
 \Line(0,14)(50,14)%
 \Line(0,16)(50,16)%
 \CCirc(17.5,15){3}{Black}{Black}
}}
\def\yprocEb{\pic{%
 \Agl(25,15)(15,5,175)%
 \Lgl(25,16)(25,29.5)%
 \Line(0,14)(50,14)%
 \Line(0,16)(50,16)%
 \CCirc(32.5,15){3}{Black}{Black}
}}
\def\yprocEc{\pic{%
 \Agl(25,15)(15,5,175)%
 \Lgl(25,16)(25,29.5)%
 \Line(0,14)(50,14)%
 \Line(0,16)(50,16)%
 \CCirc(17.5,15){3}{Black}{Black}
 \CCirc(32.5,15){3}{Black}{Black}
}}
\def\yprocFa{\pic{%
 \CArc(25,15)(15,5,175)%
 \Lgl(25,16)(25,29.5)%
 \Line(0,14)(50,14)%
 \Line(0,16)(50,16)%
 \CCirc(17.5,15){3}{Black}{Black}
}}
\def\yprocFb{\pic{%
 \CArc(25,15)(15,5,175)%
 \Lgl(25,16)(25,29.5)%
 \Line(0,14)(50,14)%
 \Line(0,16)(50,16)%
 \CCirc(32.5,15){3}{Black}{Black}
}}
\def\yprocFc{\pic{%
 \CArc(25,15)(15,5,175)%
 \Lgl(25,16)(25,29.5)%
 \Line(0,14)(50,14)%
 \Line(0,16)(50,16)%
 \CCirc(17.5,15){3}{Black}{Black}
 \CCirc(32.5,15){3}{Black}{Black}
}}
\def\yprocGa{\pic{%
 \CArc(25,15)(15,90,175)%
 \Agl(25,15)(15,5,90)%
 \Line(25,16)(25,29.5)%
 \Line(0,14)(50,14)%
 \Line(0,16)(50,16)%
 \CCirc(17.5,15){3}{Black}{Black}
}}
\def\yprocGb{\pic{%
 \CArc(25,15)(15,90,175)%
 \Agl(25,15)(15,5,90)%
 \Line(25,16)(25,29.5)%
 \Line(0,14)(50,14)%
 \Line(0,16)(50,16)%
 \CCirc(32.5,15){3}{Black}{Black}
}}
\def\yprocGc{\pic{%
 \CArc(25,15)(15,90,175)%
 \Agl(25,15)(15,5,90)%
 \Line(25,16)(25,29.5)%
 \Line(0,14)(50,14)%
 \Line(0,16)(50,16)%
 \CCirc(17.5,15){3}{Black}{Black}
 \CCirc(32.5,15){3}{Black}{Black}
}}
\def\yprocHa{\pic{%
 \Agl(25,15)(15,90,175)%
 \CArc(25,15)(15,5,90)%
 \Line(25,16)(25,29.5)%
 \Line(0,14)(50,14)%
 \Line(0,16)(50,16)%
 \CCirc(17.5,15){3}{Black}{Black}
}}
\def\yprocHb{\pic{%
 \Agl(25,15)(15,90,175)%
 \CArc(25,15)(15,5,90)%
 \Line(25,16)(25,29.5)%
 \Line(0,14)(50,14)%
 \Line(0,16)(50,16)%
 \CCirc(32.5,15){3}{Black}{Black}
}}
\def\yprocHc{\pic{%
 \Agl(25,15)(15,90,175)%
 \CArc(25,15)(15,5,90)%
 \Line(25,16)(25,29.5)%
 \Line(0,14)(50,14)%
 \Line(0,16)(50,16)%
 \CCirc(17.5,15){3}{Black}{Black}
 \CCirc(32.5,15){3}{Black}{Black}
}}
\def\yprocI{\pic{%
 \Agl(19,15)(13,5,175)%
 \Agl(31,15)(13,5,105)%
 \Agl(31,15)(13,135,175)%
 \Line(0,14)(50,14)%
 \Line(0,16)(50,16)%
}}
\def\yprocId{\pic{%
 \Agl(19,15)(13,5,175)%
 \Agl(31,15)(13,5,105)%
 \Agl(31,15)(13,135,175)%
 \Line(0,14)(50,14)%
 \Line(0,16)(50,16)%
 \CCirc(12,15){3}{Black}{Black}
 \CCirc(25,15){3}{Black}{Black}
}}
\def\yprocIf{\pic{%
 \Agl(19,15)(13,5,175)%
 \Agl(31,15)(13,5,105)%
 \Agl(31,15)(13,135,175)%
 \Line(0,14)(50,14)%
 \Line(0,16)(50,16)%
 \CCirc(25,15){3}{Black}{Black}
 \CCirc(38,15){3}{Black}{Black}
}}
\def\yprocJ{\pic{%
 \CArc(19,15)(13,5,175)%
 \Agl(31,15)(13,5,105)%
 \Agl(31,15)(13,135,175)%
 \Line(0,14)(50,14)%
 \Line(0,16)(50,16)%
}}
\def\yprocJd{\pic{%
 \CArc(19,15)(13,5,175)%
 \Agl(31,15)(13,5,105)%
 \Agl(31,15)(13,135,175)%
 \Line(0,14)(50,14)%
 \Line(0,16)(50,16)%
 \CCirc(12,15){3}{Black}{Black}
 \CCirc(25,15){3}{Black}{Black}
}}
\def\yprocJf{\pic{%
 \CArc(19,15)(13,5,175)%
 \Agl(31,15)(13,5,105)%
 \Agl(31,15)(13,135,175)%
 \Line(0,14)(50,14)%
 \Line(0,16)(50,16)%
 \CCirc(25,15){3}{Black}{Black}
 \CCirc(38,15){3}{Black}{Black}
}}
\def\yprocK{\pic{%
 \Agl(19,15)(13,5,175)%
 \CArc(31,15)(13,5,105)%
 \CArc(31,15)(13,135,175)%
 \Line(0,14)(50,14)%
 \Line(0,16)(50,16)%
}}
\def\yprocKd{\pic{%
 \Agl(19,15)(13,5,175)%
 \CArc(31,15)(13,5,105)%
 \CArc(31,15)(13,135,175)%
 \Line(0,14)(50,14)%
 \Line(0,16)(50,16)%
 \CCirc(12,15){3}{Black}{Black}
 \CCirc(25,15){3}{Black}{Black}
}}
\def\yprocKf{\pic{%
 \Agl(19,15)(13,5,175)%
 \CArc(31,15)(13,5,105)%
 \CArc(31,15)(13,135,175)%
 \Line(0,14)(50,14)%
 \Line(0,16)(50,16)%
 \CCirc(25,15){3}{Black}{Black}
 \CCirc(38,15){3}{Black}{Black}
}}
\def\yprocL{\pic{%
 \CArc(19,15)(13,5,175)%
 \CArc(31,15)(13,5,105)%
 \CArc(31,15)(13,135,175)%
 \Line(0,14)(50,14)%
 \Line(0,16)(50,16)%
}}
\def\yprocLd{\pic{%
 \CArc(19,15)(13,5,175)%
 \CArc(31,15)(13,5,105)%
 \CArc(31,15)(13,135,175)%
 \Line(0,14)(50,14)%
 \Line(0,16)(50,16)%
 \CCirc(12,15){3}{Black}{Black}
 \CCirc(25,15){3}{Black}{Black}
}}
\def\yprocLf{\pic{%
 \CArc(19,15)(13,5,175)%
 \CArc(31,15)(13,5,105)%
 \CArc(31,15)(13,135,175)%
 \Line(0,14)(50,14)%
 \Line(0,16)(50,16)%
 \CCirc(25,15){3}{Black}{Black}
 \CCirc(38,15){3}{Black}{Black}
}}
\def\yprocM{\pic{%
 \Agl(25,15)(19,5,175)%
 \Agl(25,15)(7,5,175)%
 \Line(0,14)(50,14)%
 \Line(0,16)(50,16)%
}}
\def\yprocMb{\pic{%
 \Agl(25,15)(19,5,175)%
 \Agl(25,15)(7,5,175)%
 \Line(0,14)(50,14)%
 \Line(0,16)(50,16)%
 \CCirc(25,15){3}{Black}{Black}
}}
\def\yprocMe{\pic{%
 \Agl(25,15)(19,5,175)%
 \Agl(25,15)(7,5,175)%
 \Line(0,14)(50,14)%
 \Line(0,16)(50,16)%
 \CCirc(12,15){3}{Black}{Black}
 \CCirc(38,15){3}{Black}{Black}
}}
\def\yprocMg{\pic{%
 \Agl(25,15)(19,5,175)%
 \Agl(25,15)(7,5,175)%
 \Line(0,14)(50,14)%
 \Line(0,16)(50,16)%
 \CCirc(12,15){3}{Black}{Black}
 \CCirc(25,15){3}{Black}{Black}
 \CCirc(38,15){3}{Black}{Black}
}}
\def\yprocN{\pic{%
 \Agl(25,15)(19,5,175)%
 \CArc(25,15)(7,5,175)%
 \Line(0,14)(50,14)%
 \Line(0,16)(50,16)%
}}
\def\yprocNb{\pic{%
 \Agl(25,15)(19,5,175)%
 \CArc(25,15)(7,5,175)%
 \Line(0,14)(50,14)%
 \Line(0,16)(50,16)%
 \CCirc(25,15){3}{Black}{Black}
}}
\def\yprocNe{\pic{%
 \Agl(25,15)(19,5,175)%
 \CArc(25,15)(7,5,175)%
 \Line(0,14)(50,14)%
 \Line(0,16)(50,16)%
 \CCirc(12,15){3}{Black}{Black}
 \CCirc(38,15){3}{Black}{Black}
}}
\def\yprocNg{\pic{%
 \Agl(25,15)(19,5,175)%
 \CArc(25,15)(7,5,175)%
 \Line(0,14)(50,14)%
 \Line(0,16)(50,16)%
 \CCirc(12,15){3}{Black}{Black}
 \CCirc(25,15){3}{Black}{Black}
 \CCirc(38,15){3}{Black}{Black}
}}
\def\yprocO{\pic{%
 \CArc(25,15)(19,5,175)%
 \Agl(25,15)(7,5,175)%
 \Line(0,14)(50,14)%
 \Line(0,16)(50,16)%
}}
\def\yprocOb{\pic{%
 \CArc(25,15)(19,5,175)%
 \Agl(25,15)(7,5,175)%
 \Line(0,14)(50,14)%
 \Line(0,16)(50,16)%
 \CCirc(25,15){3}{Black}{Black}
}}
\def\yprocOe{\pic{%
 \CArc(25,15)(19,5,175)%
 \Agl(25,15)(7,5,175)%
 \Line(0,14)(50,14)%
 \Line(0,16)(50,16)%
 \CCirc(12,15){3}{Black}{Black}
 \CCirc(38,15){3}{Black}{Black}
}}
\def\yprocOg{\pic{%
 \CArc(25,15)(19,5,175)%
 \Agl(25,15)(7,5,175)%
 \Line(0,14)(50,14)%
 \Line(0,16)(50,16)%
 \CCirc(12,15){3}{Black}{Black}
 \CCirc(25,15){3}{Black}{Black}
 \CCirc(38,15){3}{Black}{Black}
}}
\def\yprocP{\pic{%
 \CArc(25,15)(19,5,175)%
 \CArc(25,15)(7,5,175)%
 \Line(0,14)(50,14)%
 \Line(0,16)(50,16)%
}}
\def\yprocPb{\pic{%
 \CArc(25,15)(19,5,175)%
 \CArc(25,15)(7,5,175)%
 \Line(0,14)(50,14)%
 \Line(0,16)(50,16)%
 \CCirc(25,15){3}{Black}{Black}
}}
\def\yprocPe{\pic{%
 \CArc(25,15)(19,5,175)%
 \CArc(25,15)(7,5,175)%
 \Line(0,14)(50,14)%
 \Line(0,16)(50,16)%
 \CCirc(12,15){3}{Black}{Black}
 \CCirc(38,15){3}{Black}{Black}
}}
\def\yprocPg{\pic{%
 \CArc(25,15)(19,5,175)%
 \CArc(25,15)(7,5,175)%
 \Line(0,14)(50,14)%
 \Line(0,16)(50,16)%
 \CCirc(12,15){3}{Black}{Black}
 \CCirc(25,15){3}{Black}{Black}
 \CCirc(38,15){3}{Black}{Black}
}}
\def\selfSmaster{\pic{%
 \Lgl(0,15)(13,15)%
 \Lgl(17,15)(30,15)%
 \CCirc(15,15){5}{Black}{Gray}
}}
\def\selfA{\pic{%
 \Lgl(0,15)(8,15)%
 \Lgl(22,15)(30,15)%
 \Agl(15,15)(7,0,360)
}}
\def\selfB{\pic{%
 \Lgl(0,15)(8,15)%
 \Lgl(22,15)(30,15)%
 \Agh(15,15)(7,0,180)
 \Agh(15,15)(7,180,360)
}}
\def\selfC{\pic{%
 \Lgl(0,15)(15,13)%
 \Lgl(15,13)(30,15)%
 \Agl(15,21)(7,0,360)
}}
\def\selfD{\pic{%
 \Lgl(0,15)(8,15)%
 \Lgl(22,15)(30,15)%
 \Asc(15,15)(7,0,180)
 \Asc(15,15)(7,180,360)
}}
\def\selfE{\pic{%
 \Lgl(0,15)(15,13)%
 \Lgl(15,13)(30,15)%
 \Asc(15,21)(7,0,360)
}}
\def\selfF{\pic{%
 \Lgl(0,15)(8,15)%
 \Lgl(22,15)(30,15)%
 \Asc(15,15)(7,0,180)
 \Agl(15,15)(7,180,360)
}}
\def\selfG{\pic{%
 \Lgl(0,15)(8,15)%
 \Lgl(22,15)(30,15)%
 \Aqq(15,15)(7,0,180)
 \Aqq(15,15)(7,180,360)
}}
\def\selfH{\pic{%
 \Lgl(0,15)(15,13)%
 \Lgl(15,13)(30,15)%
 \Aqq(15,21)(7,0,360)
}}
\def\selfI{\pic{%
 \Lgl(0,15)(15,13)%
 \Lgl(15,13)(30,15)%
 \Lsc(15,13)(15,23)
 \Agl(15,30)(7,0,360)
}}
\def\selfJ{\pic{%
 \Lgl(0,15)(15,13)%
 \Lgl(15,13)(30,15)%
 \Lsc(15,13)(15,23)
 \Agh(15,30)(7,-90,270)
}}
\def\selfK{\pic{%
 \Lgl(0,15)(15,13)%
 \Lgl(15,13)(30,15)%
 \Lsc(15,13)(15,23)
 \Asc(15,30)(7,0,360)
}}
\def\selfL{\pic{%
 \Lgl(0,15)(15,13)%
 \Lgl(15,13)(30,15)%
 \Lsc(15,13)(15,23)
 \Aqq(15,30)(7,0,360)
}}
\def\selfTmaster{\pic{%
 \Lqq(0,15)(13,15)%
 \Lqq(17,15)(30,15)%
 \CCirc(15,15){5}{Black}{Gray}
}}
\def\selfTA{\pic{%
 \Lqq(0,15)(8,15)%
 \Lqq(22,15)(30,15)%
 \Agl(15,15)(7,0,180)
 \Aqq(15,15)(7,180,360)
}}
\def\selfTC{\pic{%
 \Lqq(0,15)(15,13)%
 \Lqq(15,13)(30,15)%
 \Agl(15,21)(7,0,360)
}}
\def\selfTE{\pic{%
 \Lqq(0,15)(15,13)%
 \Lqq(15,13)(30,15)%
 \Asc(15,21)(7,0,360)
}}
\def\selfTF{\pic{%
 \Lqq(0,15)(8,15)%
 \Lqq(22,15)(30,15)%
 \Asc(15,15)(7,0,180)
 \Aqq(15,15)(7,180,360)
}}
\def\selfTI{\pic{%
 \Lqq(0,15)(15,13)%
 \Lqq(15,13)(30,15)%
 \Lsc(15,13)(15,23)
 \Agl(15,30)(7,0,360)
}}
\def\selfTJ{\pic{%
 \Lqq(0,15)(15,13)%
 \Lqq(15,13)(30,15)%
 \Lsc(15,13)(15,23)
 \Agh(15,30)(7,-90,270)
}}
\def\selfTK{\pic{%
 \Lqq(0,15)(15,13)%
 \Lqq(15,13)(30,15)%
 \Lsc(15,13)(15,23)
 \Asc(15,30)(7,0,360)
}}
\def\selfTL{\pic{%
 \Lqq(0,15)(15,13)%
 \Lqq(15,13)(30,15)%
 \Lsc(15,13)(15,23)
 \Aqq(15,30)(7,0,360)
}}
\def\filledsquare{\parbox[c]{5pt}{%
 \begin{picture}(5,5)(0,0)
 \SetWidth{1.0}\SetScale{1.0}
 \GBox(0,0)(5,5){0}
 \end{picture}}\,
}

\makeatletter \@addtoreset{equation}{section} \makeatother
\renewcommand{\theequation}{\arabic{section}.\arabic{equation}}
\makeatletter
\renewcommand\section{\@startsection {section}{1}{\z@}%
                                   {-5.5ex \@plus -1ex \@minus -.2ex}
                                   {2.3ex \@plus.2ex}%
                                   {\normalfont\large\bfseries}}
\renewcommand\subsection{\@startsection{subsection}{2}{\z@}%
                                     {-3.25ex\@plus -1ex \@minus -.2ex}%
                                     {1.5ex \@plus .2ex}%
                                     {\normalfont\normalsize\bfseries}}
\renewcommand\thesection {\@arabic\c@section}
\renewcommand\thesubsection   {\thesection.\@arabic\c@subsection}
\renewcommand{\@seccntformat}[1]{%
\csname the#1\endcsname.\hspace{1.0em}}
\makeatother

\begin{document}

\flushbottom

\begin{titlepage}

\begin{flushright}
October 2019
\end{flushright}
\begin{centering}

\vfill

{\Large{\bf
  A thermal neutrino interaction rate at NLO 
}} 

\vspace{0.8cm}

G.~Jackson 
 and 
M.~Laine 
 
\vspace{0.8cm}

{\em
AEC, 
Institute for Theoretical Physics, 
University of Bern, \\ 
Sidlerstrasse 5, CH-3012 Bern, Switzerland \\}

\vspace*{0.8cm}

\mbox{\bf Abstract}
 
\end{centering}

\vspace*{0.3cm}
 
\noindent
The interaction rate of an ultrarelativistic 
active neutrino at a temperature 
below the electroweak crossover plays a role in leptogenesis scenarios
based on oscillations between active neutrinos and GeV-scale sterile
neutrinos. By making use of a Euclideanization
property of a thermal light-cone correlator, we determine the $\rmO(g)$
correction to such an interaction rate in the high-temperature limit
$\pi T \gg \mW^{ }$, finding a $\sim 15 ... 40\%$ reduction. 
For a benchmark point, this NLO correction decreases
the lepton asymmetries produced by $\sim 1$\%.  

\vfill

 
\vspace*{1cm}
  
\vfill

\end{titlepage}

%
\section{Introduction}
\la{se:intro} 

Completing the Standard Model with GeV-scale sterile 
neutrinos has become popular in recent years, given that they may 
account for the observed active neutrino mass differences 
and mixing angles through the seesaw mechanism~\cite{ss1,ss2,ss3}, 
play a role in cosmology~\cite{ars,as}, 
and be searched for experimentally
(for a review cf.,\ e.g.,\ ref.~\cite{exp}). 
Apart from generating a baryon asymmetry
(cf.,\ 
e.g.,\ refs.~\cite{scan1,scan2,scan3,scan4,scan5,scan6} 
and references therein), 
their dynamics could lead to 
the generation of lepton asymmetries larger 
than the baryon asymmetry~\cite{singlet,drewes,eijima,simultaneous}, 
which could influence late-time cosmology, 
such as dark matter production~\cite{sf,shifuller,dmpheno}. 

If right-handed neutrinos are added to the Standard Model 
in a minimal renormalizable way, without introducing any other
dynamical fields, then the neutrino sector 
is fully characterized by the values of Majorana mass
parameters and neutrino Yukawa couplings. 
Even if this amounts to a large-dimensional parameter
space, the values of the 
parameters start to be constrained. 
Therefore, it is important
to scrutinize the precision of the theoretical computations on which 
the cosmological significance of this model relies.
This is the goal of 
the present study. Specifically, we aim to determine one of the 
important rates, defined in \eq\nr{def_Sigma}, 
up to next-to-leading order (NLO) 
in the weak-coupling expansion. 

The computation of NLO corrections to real-time rates 
is a challenging task in the 
so-called ultrarelativistic regime $\pi T \gg m^{ }_i$, 
where $m^{ }_i$ refers to the masses of the plasma particles, 
hampered as it is by powerlike 
infrared divergences which lead to the breakdown of the naive loop 
expansion.\footnote{%
 There are challenges also in the  
 relativistic ($\pi T \sim m^{ }_i$) and 
 non-relativistic ($\pi T \ll m^{ }_i$) regimes
 but those are of a different nature and parametrically
 less severe, 
 cf.\ e.g.\ refs.~\cite{nlo,nlo1,nlo2,nlo3,nlo4}. 
} 
In general, a nested resummation of the loop expansion is 
necessary for generating a consistent weak-coupling expansion. 
An important breakthrough was achieved in ref.~\cite{qhat}, 
where it was realized that for many ultrarelativistic 
observables the real-time 
problem can be reduced to a static one. 
Then resummations can be implemented
in a tractable fashion, which permitted for ref.~\cite{qhat} 
to recover previous results~\cite{agz}
in a simple way and to push the computation 
one order higher in the coupling. 
This insight has subsequently inspired, for instance,  
NLO determinations of the thermal photon~\cite{photon} and 
soft dilepton~\cite{dilepton} production rates from a QCD plasma; 
an estimate of the NLO contribution to its shear 
viscosity~\cite{shear}; as well as attempts at incorporating
the contribution of soft modes on a non-perturbative 
level, through numerical simulations of an effective 
theory~\cite{nonpert,clgt,prs,onofrio}. 

The goal of the present paper is 
to adapt these techniques 
to the electroweak theory. 
Even though the SU(2) 
gauge coupling $g^{ }_2\sim \frac{2}{3}$
is smaller than the QCD one, the NLO corrections are only
suppressed by $g_2^2 T /(\pi m^{ }_i) \sim g^{ }_2/\pi$
(assuming $m^{ }_i \sim g^{ }_2 T$), 
and could come with large prefactors. 
Previously, they have been determined for so-called susceptibilities, 
relating chemical potentials to lepton asymmetries.
As they were found to be numerically 
significant~\cite{washout,sangel}, it appears well motivated
to extend the exercise to a genuine rate observable. 

%
\section{Formulation of the problem}
\la{se:setup} 

Denoting by $\mathcal{K} = (\omega,\vec{k})$ 
the four-momentum of an active neutrino propagating through 
a medium at a temperature $T$ and by $\bsl{\Sigma}$ its (advanced)
self-energy, 
the chiral nature of gauge interactions (cf.\ \eq\nr{L_M})
implies that we may write~\cite{weldon}
\be
 \re\bsl{\Sigma} \; = \; a \, \bsl{\mathcal{K}} + b \, \msl{u}
 \;, \quad
 \im\bsl{\Sigma} \; = \; \frac{1}{2}
 \Bigl( 
   \Gamma^{ }_{\!\mathcal{K}} \, \bsl{\mathcal{K}}
 +  
   \Gamma^{ }_{\!u} \msl{u} 
 \Bigr)
 \;, \la{def_Sigma}
\ee
where $u \equiv (1,\vec{0})$ denotes the four-velocity of the plasma
in the local rest frame. 
The function $a$ represents a radiative ``wave function correction'', 
whereas $b$ can be interpreted as a thermal correction
to a dispersion relation~\cite{weldon}
(the full propagator is 
$
 \propto (\bsl{\mathcal{K}}+\bsl{\Sigma})^{-1}
$). In the following, 
we are concerned with the interaction rate $\Gamma^{ }_{\!u}$. 
The rate $\Gamma^{ }_{\!\mathcal{K}}$ is relevant for the subleading
helicity-flipping active-sterile 
transitions~\cite{degenerate}, however it was found 
in ref.~\cite{degenerate}, drawing upon earlier work in the 
QCD context~\cite{review,gmt}, that only the interaction rate 
$\Gamma^{ }_{\!u}$ is susceptible to a simplified Euclideanized 
treatment {\it \`a la} ref.~\cite{qhat}. 

We assume that the neutrino is, to a good approximation, 
ultrarelativistic: $k \equiv |\vec{k}| \sim \pi T \gg M$, 
where $M = \sqrt{\omega^2 - k^2}$ denotes its virtuality. 
It interacts via weak interactions, 
\be 
 \mathcal{L}^{ }_{M} = 
 \bar{\ell}^{ }_\rmii{L} i \gamma^\mu_{ } D^{ }_\mu \ell^{ }_\rmii{L}
 \;, \quad
 D^{ }_\mu = \partial^{ }_\mu - \frac{i g^{ }_1 B^{ }_{\mu}}{2}
 - \frac{i g^{ }_2\, \sigma^a_{ }A^a_{\mu}}{2}
 \;, \quad
 \ell^{ }_\rmii{L}
 \; \equiv \; 
 \biggl(\! 
   \begin{array}{c} 
     \nu^{ }_\rmii{L} \\ e^{ }_\rmii{L} 
   \end{array}
 \!\biggr)
 \;, \la{L_M}
\ee
where $g^{ }_1$ is the hypercharge coupling, $B^{ }_\mu$ is the 
corresponding gauge potential, $\sigma^a_{ }$ are the Pauli matrices,
$g^{ }_2$ is the weak coupling  
and $A^a_\mu$ are the SU$^{ }_\rmii{L}$(2) gauge potentials. 

In order to isolate the relevant helicity components, we employ
the Weyl representation of the Dirac matrices: 
$
 \gamma^{0}_{ }\gamma^{i}_{ }= \diag(-\sigma^{i}_{ },\sigma^{i}_{ }), 
$
$\aL = \diag(\mathbbm{1},0)$.
Going to momentum space, 
$\partial^{ }_\mu \to i \mathcal{K}^{ }_\mu$, and  
aligning the momentum in the $z$-direction, the free part 
$
 \mathcal{L}^{ }_M \supset 
 \nu^\dagger_\rmii{L} 
 (- \omega \mathbbm{1} - k^z_{ }\sigma^{z}_{ }) \nu^{ }_\rmii{L}
 + 
 e^\dagger_\rmii{L} 
 (- \omega \mathbbm{1} - k^z_{ }\sigma^{z}_{ }) e^{ }_\rmii{L}
$
implies that the lower (negative-helicity) components
of $\nu^{ }_\rmii{L}$ and $e^{ }_\rmii{L}$ 
go on-shell for $\omega = k^z_{ }$. 
In the following we denote this component 
by $\psi$ for $\nu^{ }_\rmii{L}$, and by $\chi$
for $e^{ }_\rmii{L}$. For these
components, the coefficients
of \eq\nr{def_Sigma} appear in the effective action as 
\ba
 \mathcal{S}^{ }_{M,\rmi{eff}} & \supset & 
 \int^{ }_{\mathcal{K}} 
 \psi^\dagger_{ }(\mathcal{K})
 \, \biggl[ 
 \bigl(-\omega + k^{z}_{ } \bigr) 
 \biggl( 
        1 + a + \frac{i \Gamma^{ }_{\!\mathcal{K}} }{2}
 \biggr) 
  - 
 \biggl( 
        b +  \frac{i \Gamma^{ }_{\! u} }{2}  
 \biggr)
 \biggr] \, 
 \psi(\mathcal{K})
 \nn 
 & + & 
 \int^{ }_{\mathcal{K}}  
 \chi^\dagger_{ }(\mathcal{K})
 \, \biggl[ 
 \bigl(-\omega + k^{z}_{ } \bigr) 
 \biggl( 
        1 + \tilde a + \frac{i \tilde\Gamma^{ }_{\!\mathcal{K}} }{2}
 \biggr) 
  - 
 \biggl( 
        \tilde b +  \frac{i \tilde \Gamma^{ }_{\! u} }{2}  
 \biggr)
 \biggr] \, 
 \chi(\mathcal{K})
 \;, \la{SM_eff}
\ea
where 
$
 \tilde{a},\tilde{b},\tilde\Gamma^{ }_{\!\mathcal{K}},\tilde \Gamma^{ }_{\! u}
$ 
refer to the properties of left-handed electrons. 

Following ref.~\cite{qhat}, the idea now is to expand in fluctuations
around the on-shell point, and then to rotate the light-like
propagation into a static one. 
Simultaneously, we go over to Euclidean conventions, 
{\it viz.} $A^\M_0 = i A^\E_0$, and define a Euclidean Lagrangian
as $L^{ }_\E \equiv - \mathcal{L}^{ }_\M$. Linear combinations
of gauge potentials are denoted by 
$
 W^{\pm}_{\mu} \;\equiv\; (A^1_\mu \mp i A^2_\mu) / \sqrt{2}
$, 
$
 \tilde{g} Z^{ }_{\mu} \;\equiv\; 
 g^{ }_1 B^{ }_\mu + g^{ }_2 A^3_\mu 
$, 
$
 \tilde{g} Z'_{\mu} \;\equiv\; 
g^{ }_1 B^{ }_\mu - g^{ }_2 A^3_\mu 
$, 
where 
$
 \tilde{g} \; \equiv \; \sqrt{g_1^2 + g_2^2}
$. 
Then static fluctuations around the 
on-shell point are described by 
\be
 L^{ }_\E \; = \;  
 - \, 
 \biggl(\! 
   \begin{array}{c} \psi \\ \chi 
   \end{array}
 \!\biggr)^\dagger_{ }
 \biggl\{ 
   \begin{array}{cc}
      i \partial^{ }_z 
     + \frac{\tilde{g}}{2} \bigl( i Z^{ }_{0} + Z^{ }_{3} \bigr)  
     &
      \frac{g^{ }_2}{\sqrt{2}} \bigl(i W^+_{0} + W^+_{3} \bigr)
     \\ 
      \frac{g^{ }_2}{\sqrt{2}} \bigl(i W^-_{0} + W^-_{3} \bigr)
     & 
      i \partial^{ }_z 
     + \frac{\tilde{g}}{2} \bigl( i Z'_{0} + Z'_{3} \bigr)  
   \end{array}
 \biggr\} 
 \biggl(\! 
   \begin{array}{c} \psi \\ \chi 
   \end{array} 
 \!\biggr)
 \;. \la{L_E}
\ee
The ``large'' $\omega$ and
$k^z_{ }$ have cancelled against each other, so the ``residual''
momentum generated by $i \partial^{ }_z$ in \eq\nr{L_E} can be
taken to be small (it is denoted by $k^{ }_z$ and is 
$\sim\tilde{g}^2 T$). 

With the Lagrangian of \eq\nr{L_E}, 
we compute the Euclidean version of \eq\nr{SM_eff}.
For nearly on-shell neutrinos,
it takes the form\footnote{%
 The original $\psi$ has been scaled by a factor $T^{1/2}$ so that 
 $\psi(\vec{x})$ has the dimension GeV and $\psi(\vec{k})$ the dimension
 GeV$^{-2}$. The same dimensions apply to the gauge potentials 
 $A^a_\mu(\vec{x})$ and $A^a_\mu(\vec{k})$, respectively. 
 }
\be
 S^{ }_{\E,\rmi{eff}} \supset \int_\vec{k} \psi^\dagger(\vec{k}) 
 \, \bigl[ 
   k^{ }_z + \Sigma(k^{ }_z) 
 \bigr] \,
 \psi(\vec{k}) 
 \;, \quad
 \Sigma(k^{ }_z) = 
 k^{ }_z\, 
 \biggl( 
        a + \frac{i \Gamma^{ }_{\!\mathcal{K}} }{2}
 \biggr) 
  +
        b +  \frac{i \Gamma^{ }_{\! u} }{2}  
  + \rmO(k_z^2)
 \;. \la{Sigma}
\ee
It turns out that within the effective theory, 
the real part of $\Sigma$ is odd in $k^{ }_z$ and the 
imaginary part is even in $k^{ }_z$, guaranteeing that 
correlations decay exponentially (i.e.\ that the pole is
on the imaginary axis). Consequently, only $a$ and 
$\Gamma^{ }_{\! u}$ are generated within our computation, 
as already alluded to above. 

It can be deduced from \eq\nr{Sigma} 
that the free neutrino and electron propagators have the forms 
\be
 \langle \psi(\vec{k}) \psi^\dagger_{ }(\vec{q})\rangle^{ }_0
 = 
 \frac{(2\pi)^d\delta^{(d)}(\vec{k-q})}{k_z^{ } + i 0^+}
 =
 \langle \chi(\vec{k}) \chi^\dagger_{ }(\vec{q})\rangle^{ }_0
 \;, \la{psi_prop} 
\ee
where $d = 3 -2\epsilon$. 
The other propagators are those of the 
dimensionally reduced theory for the Standard Model~\cite{generic},
with the temporal gauge field components $A^a_0$ and $B^{ }_0$
kept as dynamical fields. 
We carry out computations in a general $R^{ }_\xi$ gauge, whereby the spatial $W^{\pm}_{ }$ propagator reads
($ a',b' \in \{ 1,2 \} $)
\be
 \Bigl\langle A^{a'}_{i}\!(\vec{k}) A^{b'}_{j}\!(\vec{q})
 \Bigr\rangle^{ }_{0}
 \; = \; 
 \delta_{ }^{a'b'} T \, (2\pi)^d_{ }\delta^{(d)}_{ }(\vec{k+q})
 \biggl\{ 
  \frac{\delta^{ }_{ij}}{k^2 + \mW^2}
  + \frac{k^{ }_i k^{ }_j}{\mW^2}
  \biggl(
   \frac{1}{k^2 + \mW^2} - \frac{1}{k^2 + \xi \mW^2} 
  \biggr)
 \biggr\} 
 \;.  \la{W_prop}
\ee
There are similar propagators for the neutral components
$Z^{ }_i$ and $Q^{ }_i$ which, as usual, are obtained from 
$A^3_i$ and~$B^{ }_i$ by a rotation with the mixing angle 
$\sin(2\theta) = 2 g^{ }_1 g^{ }_2 / (g_1^2 + g_2^2)$.

For the temporal components $Z^{ }_0$ and $Q^{ }_0$, 
the mixing is modified by thermal (Debye) masses.
Let us denote the mixing angle of 
the temporal components by $\tilde\theta$.
Following ref.~\cite{broken}, 
the masses of the diagonalized modes are denoted by 
$\mZt^2$ and 
$\mQt^2$. The original gauge fields
can be expressed in the new basis as
\be
 Z^{ }_0   =   \cos(\theta - \tilde\theta)\, \widetilde{Z}^{ }_0 
               + \sin(\theta - \tilde\theta)\, \widetilde{Q}^{ }_0 
 \;, \quad
 Z'_0   =   - \cos(\theta + \tilde\theta)\, \widetilde{Z}^{ }_0 
               + \sin(\theta + \tilde\theta)\, \widetilde{Q}^{ }_0 
 \;, 
\ee
and the corresponding propagators take the forms
\ba
 \bigl\langle W^{+}_{0}(\vec{k}) W^{-}_{0}(\vec{q}) \bigr\rangle^{ }_{0}
 & = & 
 T
 (2\pi)^d_{ }\delta^{(d)}_{ }(\vec{k+q}) \, 
 \frac{1}{k^2 + \mWt^2}
 \;, \la{Wprop} \\ 
 \bigl\langle Z^{ }_{0}(\vec{k}) Z^{ }_{0}(\vec{q}) \bigr\rangle^{ }_{0}
 & = & 
 T
 (2\pi)^d_{ }\delta^{(d)}_{ }(\vec{k+q}) \, 
 \biggl[ 
 \frac{\cos^2(\theta-\tilde\theta)}{k^2 + \mZt^2}
 + 
 \frac{\sin^2(\theta-\tilde\theta)}{k^2 + \mQt^2}
 \biggr]
 \;, \la{Zprop} \\ 
 \bigl\langle Z'_{0}(\vec{k}) Z'_{0}(\vec{q}) \bigr\rangle^{ }_{0}
 & = &
 T
 (2\pi)^d_{ }\delta^{(d)}_{ }(\vec{k+q}) \, 
 \biggl[ 
 \frac{\cos^2(\theta+\tilde\theta)}{k^2 + \mZt^2}
 + 
 \frac{\sin^2(\theta+\tilde\theta)}{k^2 + \mQt^2}
 \biggr]
 \;.
\ea
The relations between 
$\mWt^{2}$, $\mZt^{2}$, $\mQt^{2}$ and the angles
$\theta,\tilde\theta$ satisfy identities which are not always easy 
to recognize at first sight, e.g.\ 
\be
 \mZt^2 - \mWt^2  \; = \; 
 \mW^2 \tan(\theta)\tan(\tilde\theta)
 \;,\quad
 \mWt^2 - \mQt^2  \; = \; 
 \mW^2 \tan(\theta)\cot(\tilde\theta)
 \;.
\ee

Let us end this section by commenting upon differences with respect to the
observable analyzed in the QCD context~\cite{qhat}. 
The quantity considered there was a Wilson loop rather than a single
propagator (which can be interpreted as a Wilson line). The 
transverse coordinate of the loop, ${r}^{ }_\perp$, served to define
a ``collision kernel'', $C(q^{ }_\perp)$, 
and the so-called jet quenching parameter, $\hat{q}$, 
is a weighted integral over $C(q^{ }_\perp)$. The weighting
by $q_\perp^2$ implies
that the soft contribution to $\hat{q}$ is UV 
divergent. In our case there is no such weighting, and $\Gamma^{ }_{\!u}$
is UV finite. On the other hand, 
the absence of weighting makes $\Gamma^{ }_{\!u}$ 
more IR sensitive than $\hat{q}$, and therefore
$\Gamma^{ }_{\!u}$ is perturbatively computable only in the Higgs phase. 

%
\section{Leading-order computation}
\la{se:lo} 

%
\begin{figure}[t]
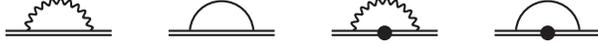


\begin{eqnarray*}
&& 
 \hspace*{-1.0cm}
 \xprocA
 \hspace*{0.90cm}
 \xprocB
 \hspace*{0.90cm}
 \xprocC
 \hspace*{0.90cm}
 \xprocD
\end{eqnarray*}


\caption[a]{\small 
 Leading-order contributions to self-energy. 
 A double line denotes a neutrino ($\psi$) propagator
 from \eq\nr{psi_prop}; a filled blob an 
 electron ($\chi$) propagator; a wiggly line a spatial gauge field; and a solid line a
 temporal gauge field, 
 which in the effective theory has turned into
 an adjoint scalar field. 
} 
\la{fig:lo}
\end{figure}
%

The leading contribution to $\Gamma^{ }_{\!u}$ originates at 1-loop 
level, through the graphs shown in \fig\ref{fig:lo}. 

Inserting the propagators from \eqs\nr{psi_prop}, \nr{W_prop}, 
\nr{Wprop} and \nr{Zprop} and 
denoting the internal momentum by 
$\vec{p} = \vec{p}^{ }_\perp + p^{ }_z \vec{e}^{ }_z$, 
we are faced with integrals of the type 
\be
 I(k_z^{ }) = T \int_\vec{p} \frac{1}{k_z^{ } - p_z^{ } + i 0^+_{ }} 
 \frac{1}{ p_z^2 + \epsilon_p^2}
 \;, \quad
 \epsilon_p^2 \;\equiv\; p_\perp^2 + m^2
 \;. 
\ee 
Noting that the external momentum 
$k_z^{ }\sim\tilde{g}^2T$ is small compared
with the mass scales $m^{ }_i \sim \tilde{g}T$, 
we can expand in $k^{ }_z$. 
Evaluating the integral over $p_z^{ }$ with the residue theorem, or with
\be
 \frac{1}{p^{ }_z - k^{ }_z - i 0^+_{ }}
 = \mathbbm{P}\biggl( \frac{1}{p^{ }_z - k^{ }_z} \biggr)
 + i \pi \delta(p^{ }_z - k^{ }_z)
 \;, \la{master}
\ee
where $\mathbbm{P}$ denotes a principal value, 
yields
\be
 I(k^{ }_z) \approx 
 T \, \biggl\{ \, 
 \frac{i}{2}
 \int_{\vec{p}_\perp} \frac{1}{\epsilon_p^2}
 + 
 k^{ }_z \, \int_\vec{p} 
 \mathbbm{P}\biggl[ 
   \frac{1}{p_z^2(p_z^2 + \epsilon_p^2)}
 \biggr] 
 + \rmO(k_z^2)
 \; \biggr\} 
 \;. \la{I}
\ee
The first term contributes directly to the width, 
whereas the second one
gives a wave function correction, proportional
to $k^{ }_z$ (cf.\ \eq\nr{Sigma}).
The wave function corrections
are gauge dependent and IR-sensitive, but they do play a role in 
cancelling similar effects from loop diagrams. 
Therefore they need to be accounted for at NLO, 
as discussed in appendix~\ref{ss:wave_fcn}.

Focussing now on the leading-order (LO) width, 
which originates from the first term
in \eq\nr{I}, it is straightforward 
to verify that there is no gauge parameter dependence.
Summing together the graphs, we obtain
($\D^{ }_i \equiv (p_\perp^2 + m_i^2 )^{-1}$)
\be
 \Gamma^\rmii{(LO)}_{\!u}
 = 
 \frac{\tilde{g}^2_{ } T}{4} \int_{\vec{p}_\perp}
 \Bigl\{
     \D_\Z^{ }
   - \cos^2(\theta-\tilde\theta) \D_\Zt^{ }
   - \sin^2(\theta-\tilde\theta) \D_\Qt^{ }
   + 
    2 \cos^2(\theta) 
   \bigl(
     \D_\W^{ } - \D_\Wt^{ }
   \bigr)
 \Bigr\} 
 \;. \la{lo_result} 
\ee
 
It remains to carry out the integral over 
the transverse momentum $\vec{p}^{ }_\perp$. 
Even if the whole integral is UV finite, 
it was argued in ref.~\cite{degenerate} that 
it is reasonable to adopt a phenomenological cutoff
$|\vec{p}^{ }_\perp| \le 2 k \sim 2\pi T$. 
Adopting this prescription, which amounts
to a partial inclusion of higher order contributions in an expansion
in $\sim m_i^2/k^2$, transverse integrals evaluate to
\be
 \int_{|\vec{p}^{ }_\perp| \le 2 k} 
 \frac{1}{p_\perp^2 + m^2}
 \; = \; 
 \frac{  \ln\bigl( 1 + {4 k^2}/{m^2} \bigr)  }{4\pi}
 \;. \la{p_perp_int}
\ee
Thereby \eq\nr{lo_result} reproduces the 
LO result for $\Gamma^{ }_{\!u}$ as 
given in \eq(5.23) of ref.~\cite{degenerate}.

%
\section{NLO result and some of its features}
\la{se:nlo} 

Proceeding to the NLO level, there are a number of contributions, 
listed in appendix~A. 
Summing them together, the result can be expressed as 
\ba
 && \hspace*{-1.5cm}
 \Gamma_u^\rmii{(NLO)} \; = \; 
 \frac{\tilde{g}^4 T^2}{8}
 \lim_{\mQ^{ }\to 0}
 \int_{\vec{p}^{ }_\perp}
 \biggl\{ 
  \sum_i 
  \big(\, c^{ }_i \A{i} + \dot c^{ }_i \Ad{i} \,\big) 
  + 
\sum_{i,j} \big( \, c^{ }_{ij} \B{i}{j}
                      + c^{\T}_{ij} \BT{i}{j}
                      + \dot c^{ }_{ij} \Bd{i}{j} \, \big)
 \biggr\}
 \;, \la{nlo_result}
\ea
where $c^{ }_i,\dot c^{ }_i, c^{ }_{ij}, c_{ij}^{\T}, \dot c^{ }_{ij}\,$ 
are coefficients
and $i,j = \{ H, Z, \ldots \}$ label particles appearing in the loops.
The ``photon mass'' $\mQ^{ }$ was introduced
as an intermediate IR regulator (see below). 
Making use of the notation of 
appendix~\ref{ss:masters}, simplified further by denoting 
$\A{i} \equiv A(m^{ }_i)$, etc, 
the linear combinations 
needed in \eq\nr{nlo_result} read\footnote{%
 For convenience a {\it c}-program evaluating this expression
 is included as an ancillary file in the arXiv record. 
 }
\ba
 && \hspace*{-0.5cm} \sum_i c^{ }_i \A{i} \ = \
\label{ci}
\\ & & 
 \Big\{   \A{H} + \frac{\mZ^2}{\mH^2} 
  \big[ 
    (d-1) \A{Z} + 
    \cos^2(\theta - \tilde\theta) \A{\wv Z} +
    \sin^2(\theta - \tilde\theta) \A{\wv Q} 
  \big]
  + \frac{2\mW^2}{\mH^2} 
    \big[\, 
      (d-1) \A{W} +  \A{\wv W}
    \,\big] 
 \Big\}
 \nn & & \; \times \,     
 \Big\{ \ 
  \D_\Z^2  - 
  \bigl[
      \cos^2(\theta - \tilde\theta) \D_\Zt^{ } +
      \sin^2(\theta - \tilde\theta) \D_\Qt^{ } \bigr]^2 + 
      2 \cos^4 (\theta) \big( \D_\W^2 - \D_\Wt^2 
     \bigr)
\ \Big\}
 \nn & - & %
  4 \cos^4 (\theta)  \, 
 \big[ 
  (d-2)\cos^2 (\theta )  \A{Z} +
  \cos^2 (\tilde\theta)  \A{\wv Z} + 
  \sin^2 (\tilde\theta)  \A{\wv Q} 
  + (d-2) \A{W} + \A{\wv W} 
 \big]
 \big( \D_\W^2 - \D_\Wt^2 \big)
 \nn  & + &
  4 \cos^2 (\theta)  
 \, \big[ 
    (d-2) \A{W} + \A{\wv W} 
    \big] 
 \nn & & \; \times \, 
 \big\{  
  \big[
    \cos (\tilde\theta) \cos(\theta - \tilde\theta) \D_\Zt^{ } -
    \sin (\tilde\theta) \sin(\theta - \tilde\theta) \D_\Qt^{ } 
  \big]^2
  - \cos^2 (\theta) \, \D_\Z^2
 \big\}
 \nn & + & 
 \A{W} \sin (2\theta)  
 \, \big\{ 
  \sin(2\theta) \D_\Z^2 + 
  \sin(2\tilde\theta) 
  \bigl[ 
   \sin^2(\theta-\tilde\theta) \D_\Qt^2 -
   \cos^2(\theta-\tilde\theta) \D_\Zt^2 
  \bigr]
 \nn & & \; - \,
   \cos (2\tilde\theta) \sin[2(\theta-\tilde\theta)] \D_\Qt^{ } \D_\Zt^{ }
 \bigr\} 
 \;, 
 \nn[2mm] 
 && \hspace*{-0.5cm} \sum_{i} \dot c^{ }_{i} \Ad{i} \ = \
 \label{cidot}
 \\ & & 
 8 \cos^4 (\theta) \, 
 \big[
  \cos^2 (\theta)  \Ad{Z} -
  \cos^2 (\tilde\theta)  \Ad{\wv Z} +
  \sin^2 (\theta)  \Ad{Q} -
  \sin^2 (\tilde\theta)  \Ad{\wv Q}
  + \Ad{W}
  - \Ad{\wv W}
 \big] 
 \big(
  \D_\Wt^{ } - \D_\W^{ } 
 \big)
 \nn & + & 
    8 \cos^3 (\theta) \, 
  \big( \,
   \Ad{W} -
   \Ad{\wv W}
   \,\big)
 \big[\, 
  \cos (\tilde\theta) \cos(\theta-\tilde\theta) \D_\Zt^{ } -
  \sin (\tilde\theta) \sin(\theta-\tilde\theta) \D_\Qt^{ } -
  \cos (\theta)  \D_\Z^{ } 
 \, \big] 
 \;,
 \nn[2mm]
 && \hspace*{-0.5cm} \sum_{i,j} c^{ }_{ij} \B{i}{j} \ = \
 \label{cij}
 \\ &&  
  2 \mZ^2\, \B{H}{Z}\, \D_\Z^2
 \nn & - & 
 2 \mZ^2 
 \big[ 
  \cos^2(\theta-\tilde\theta) \B{H}{\wv Z} +
  \sin^2 (\theta-\tilde\theta) \B{H}{\wv Q} 
 \big] 
 \big[
  \cos^2(\theta - \tilde\theta) \D_\Zt^{ } +
  \sin^2(\theta - \tilde\theta) \D_\Qt^{ } 
 \big]^2
\nn & + & 
 4 \mW^2 \cos^4 (\theta) 
 \big[\ 
  \big( \B{H}{W} + \B{Z}{W} \big)\, \D_\W^2 
  - \big( \B{H}{ \wv W} + \B{Z}{\wv W} \big)\, \D_\Wt^2 
 \ \big]
\nn & - & 
 4 \mW^2 \cos^2 (\theta) \,
 \big[ 
    \cos^2(\theta+\tilde\theta) \B{\wv Z}{ W} 
  + \sin^2(\theta+\tilde\theta) \B{\wv Q}{ W} 
 \big]\, \D_\Wt^2
\nn & + & 
 16 \cos^4 (\theta) 
 \Big\{ \ 
  p_\perp^2 \,
  \Big[ \,
       \cos^2 (\theta) \, \big( \B{Z}{W} \D_\W^2 - \B{Z}{\wv W} \D_\Wt^2 \big) 
    +\ \sin^2 (\theta) \, \big( \B{Q}{W} \D_\W^2 - \B{Q}{\wv W} \D_\Wt^2 \big) 
   \nn & & 
   \; -\  \bigl[ \cos^2 (\tilde\theta)  \B{\wv Z}{W}
             +   \sin^2 (\tilde\theta)  \B{\wv Q}{W}
       \bigr] \, \D_\Wt^2
  \Big]
  +\ \bigl[  \cos^2 (\tilde\theta)  \B{\wv Z}{\wv W}
        +    \sin^2 (\tilde\theta)  \B{\wv Q}{\wv W}
     \bigr]  \, \D_\W^{ }
\  \Big\}
\nn & - & 
  4 \mW^2 ( 4\cos^2 \theta - 1 ) 
 \bigl[
  \B{W}{W} \D_\Z^2 + \cos^2 (\theta)\, \B{Z}{W}\, \D_\W^2 
 \bigr]
 + 8 \cos^4 (\theta) \, 
 \bigl( \B{W}{W} + \B{\wv W}{\wv W} \bigr) \D_\Z^{ }
\nn & + & 
 \frac{4 \B{W}{\wv W} }{\mZ^2}\, 
  \Big\{ \ 
  \cos^2 (\theta)
  - 2 \mW^2 \bigl[ 
    \cos^2 (\theta-\tilde\theta) \, \D_\Zt^{ }
  + \sin^2 (\theta-\tilde\theta) \, \D_\Qt^{ } \bigr] 
 \nn & & \; - \,
    \bigl[ 
    (\mW^2-\mWt^2)^2 + 2 p_\perp^2 (\mW^2 + \mWt^2) + p_\perp^4 
    \bigr]
 \nn & & \; \times \, 
    \bigl[ \cos (\theta-\tilde\theta) \cos (\tilde\theta) \, \D_\Zt^{ }
         - \sin (\theta-\tilde\theta) \sin (\tilde\theta) \, \D_\Qt^{ }
    \bigr]^2
 \ \Big\}
 \;,
 \nn[2mm] 
 && \hspace*{-0.5cm} \sum_{i,j} c^\T_{ij} \BT{i}{j} \ = \
 \label{cTij}
 \\ && 
 2 \big\{
  \BT{H}{Z} 
  + \bigl[ 4(d-2) \cos^4 (\theta) + \cos^2(2\theta) \bigr] \BT{W}{W}
  + 4 \cos^4 (\theta) \, \BT{\wv W}{\wv W} 
 \big\}\, \D_\Z^2
 \nn & + & 
 4 \cos^4 (\theta) \,
 \big\{\,
  \BT{H}{W} 
  + \bigl[ 4(d-2) \cos^2 \theta + 1 \bigr] \BT{Z}{W} 
  + 4 \cos^2 (\tilde\theta) \, \BT{\wv Z}{\wv W}
  \nn & & 
  \; + \, 4(d-2) \sin^2 (\theta) \, \BT{Q}{W}
  + 4 \sin^2 (\tilde\theta) \, \BT{\wv Q}{\wv W}
 \,\big\}\, \D_\W^2
 \;,
 \nn[2mm] 
 && \hspace*{-0.5cm} \sum_{i,j} \dot c^{ }_{ij} \Bd{i}{j}  \, = \,  
 \label{cijdot}
 \\ 
 && 
 8 \cos^4 (\theta) \, 
 \big\{ \
  \cos^2 (\theta) \, \big( \Bd{Z}{W} - \Bd{Z}{\wv W} \big)
  + \cos^2 (\tilde\theta) \, \big( \Bd{\wv Z}{\wv W} - \Bd{\wv Z}{W} \big)
  \nn & + & 
    \sin^2 (\theta) \, \big( \Bd{Q}{W} - \Bd{Q}{\wv W} \big)
  + \sin^2 (\tilde\theta) \, \big( \Bd{\wv Q}{\wv W} - \Bd{\wv Q}{W} \big)
  + \tfr12 \big( \Bd{W}{W}
  + \Bd{\wv W}{\wv W} \big)
  - \Bd{\wv W}{W}
 \ \big\}
 \;. 
 \nonumber 
\ea
The final integral over $\vec{p}^{ }_\perp$ is best performed 
numerically, within the domain of \eq\nr{p_perp_int}.

Our result for 
$
 \Gamma_u^\rmii{(NLO)}
$
passes a number of crosschecks. 
A simple one is that all 
gauge dependent pieces have cancelled. 
A less trivial test can be obtained
by looking at the value of the integrand 
at $p_\perp^2 \gg m_i^2$. The leading term of the Taylor expansion, 
obtained by inserting \eq\nr{BT_rel} as well as
the asymptotics from \eqs\nr{B_asymp} and \nr{dB_asymp},  
comes from the part 
$
 \sum_{i,j} c^\T_{ij} \BT{i}{j}
$
in \eq\nr{cTij}. 
Here the different pieces add up, 
whereas all other structures contain a cancellation 
between spatial and temporal contributions. 
This yields
\ba
 \Gamma_u^\rmii{(NLO)} & \supset & 
 \frac{\tilde{g}^4 T^2}{8}
 \int_{|\vec{p}^{ }_\perp| \,\gg\, m^{ }_i}
 \frac{4 [3(2d-1)\cos^4(\theta) + \sin^4(\theta)] \BT{0}{0}}{p_\perp^4}
 \nn 
 & \stackrel{d=3}{=} & 
 - \frac{\tilde{g}^4 T^2}{8}
 \int_{|\vec{p}^{ }_\perp| \,\gg\, m^{ }_i}
 \frac{15 \cos^4(\theta) + \sin^4(\theta)}{16 p_\perp^3}
 \;, \la{nlo_asymp}
\ea
where on the first line $\BT{0}{0}$ indicates that masses can be put to zero. 
This agrees perfectly with the NLO part of 
\eq(D.6) of ref.~\cite{broken}, which was obtained
by considering the IR asymptotics of a fully relativistic
but unresummed computation. 

It is also good to check that the resummed
expression is IR finite.
As charged particles appear in the loops, 
massless photons do affect the NLO 
result. Some of their contributions are  
unproblematic, but there are certain terms
where it needs to be verified
that the limit $\mQ^{ }\to 0$ can indeed be taken
in \eq\nr{nlo_result}. 
For this we note that 
\ba
 \lim_{\mQ^{ }\to 0}
 \biggl[ 
   \dot{B}(\mQ^{ },m)  
 - \frac{ \dot{A}(\mQ^{ }) }{ p^2 + m^2 }
 \biggr]
 \;\stackrel{d=3}{=}\;
 \frac{m^2 - p^2}{8\pi m(p^2+m^2)^2}
 \;.
\ea
Indeed all $\mQ^{ }$-dependence
of \eqs\nr{cidot} and \nr{cijdot} 
appears in this IR-safe combination. 

\begin{figure}[t]

\hspace*{-0.1cm}
\centerline{%
 \epsfxsize=7.4cm\epsfbox{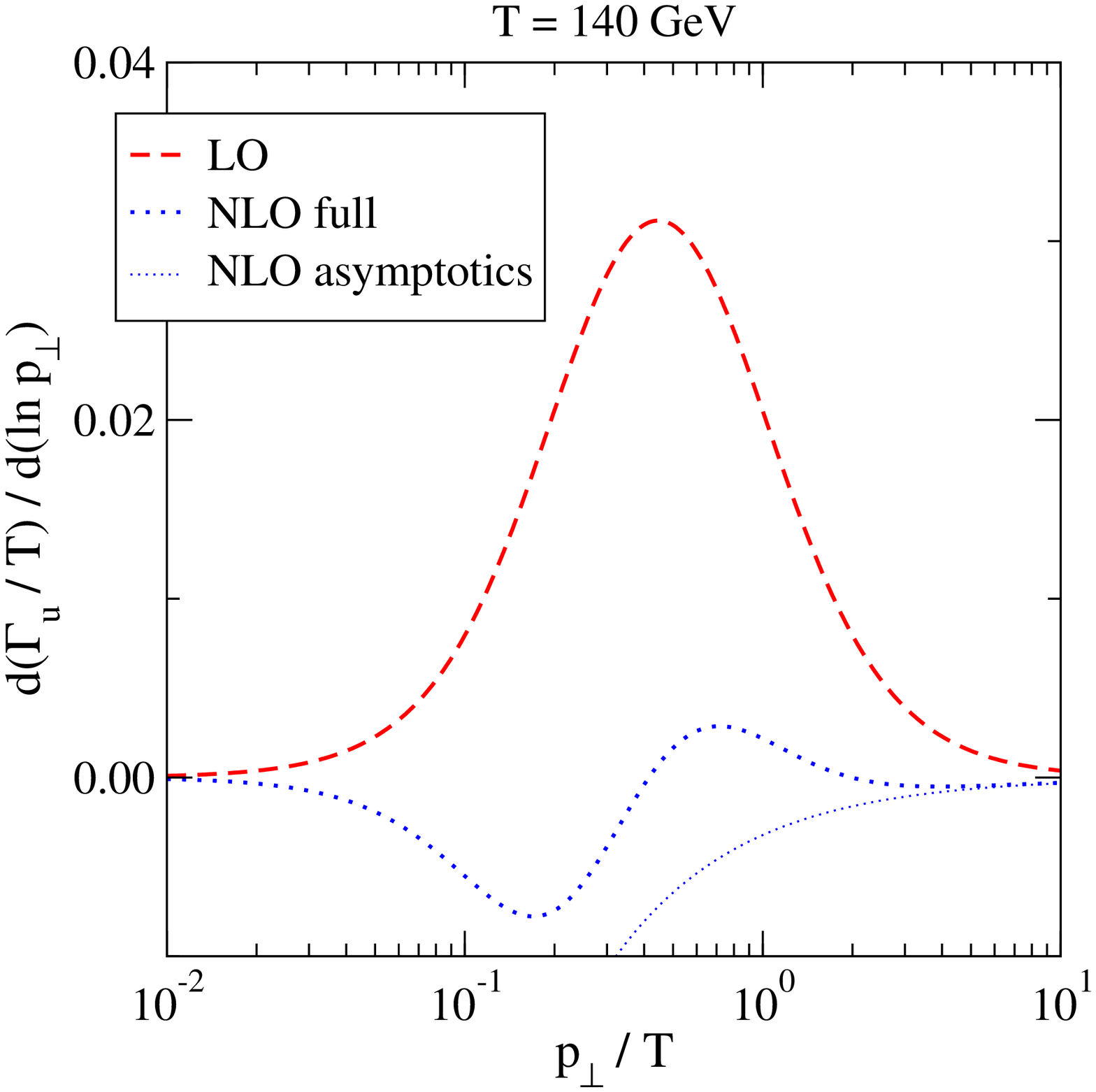}%
 \hspace*{7mm}\epsfxsize=7.6cm\epsfbox{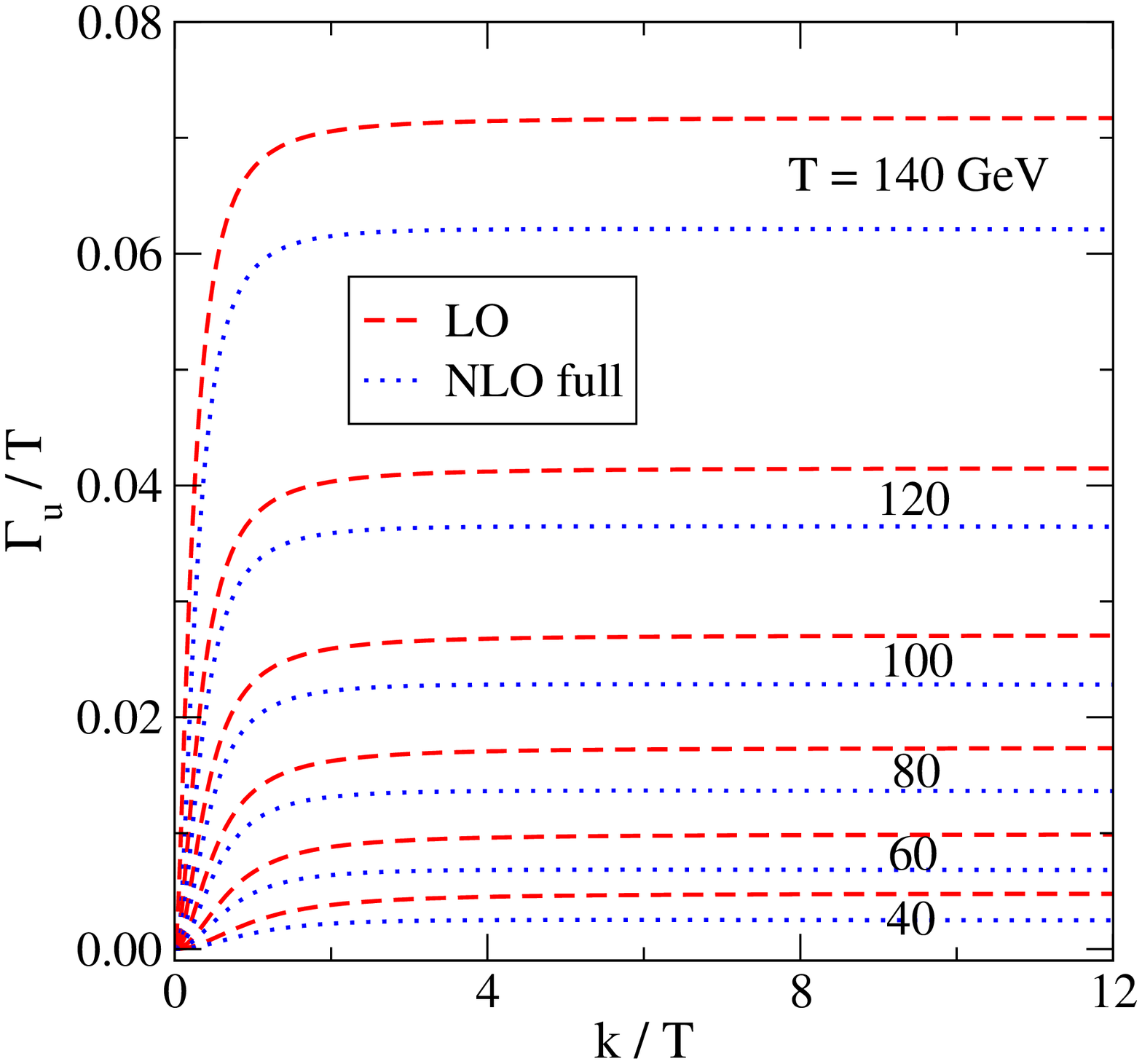}
}

\caption[a]{\small
 Left: the integrands from \eqs\nr{lo_result} (LO),
 \nr{nlo_result} (NLO full) 
 and 
 \nr{nlo_asymp} (NLO asymptotics), 
 at $T \approx 140$~GeV. 
 Right: the corresponding integrals, as a function of $T$ and $k$
 (the latter originates as discussed around \eq\nr{p_perp_int}).
 At high $T$ the NLO correction is seen to reduce the LO rate by $\sim 15\%$, 
 at low $T$ by $\sim 40\%$.
}

\la{fig:Gamma}
\end{figure}

We plot the result from \eq\nr{nlo_result}, 
both as an integrand and after the integration, in \fig\ref{fig:Gamma}.
For this, the parameters have been fixed as in ref.~\cite{degenerate}. 
A partial cancellation between contributions from various domains 
of $p^{ }_\perp$ can be observed, as a result of which the final
magnitude of the negative NLO correction remains at a modest $\sim 15\%$ level
at high temperatures. It reaches $\sim 40\%$ at low temperatures,  
mostly because the LO contribution is ``anomalously small'' there due to 
a cancellation between the spatial and temporal contributions 
in \eq\nr{lo_result}, {\it viz.}
$
  \Gamma^\rmii{(LO)}_{\!u} \sim \tilde{g}^4 T^3 / m_i^2
$
at 
$
 \tilde{g} T \ll m^{ }_i, 
$
however in that region the approximation $m^{ }_i \ll \pi T$
inherent to our effective theory approach gradually breaks down.

%
\section{Summary and outlook}
\la{se:concl} 

We have reported a thermal
NLO computation of a neutrino interaction 
rate $\Gamma^{ }_{\! u}$, 
defined through \eq\nr{def_Sigma}, 
in the temperature range $\pi T \gg \mW^{ }$. 
The final 
result is given in \eq\nr{nlo_result} and plotted numerically
in \fig\ref{fig:Gamma}, whereas many 
technical details are relegated into appendix~A. 
The result passes a number of crosschecks, 
in particular it is gauge independent and displays the UV asymptotics
predicted by an earlier unresummed computation. 
By taking $g^{ }_1\to 0$, it can partially also be contrasted with
the QCD result of ref.~\cite{qhat}, even if the presence of a Higgs
field and Higgs expectation value prohibit an unambiguous comparison. 

The coefficient $\Gamma^{ }_{\! u}$ is but one of a set of mass 
corrections and interaction rates entering a complete GeV-scale
leptogenesis framework~\cite{degenerate}.\footnote{%
 More precisely, $\Gamma^{ }_{\! u}$ determines the rate of 
 helicity-conserving active-sterile oscillations, as 
 \be
 \Gamma \simeq 
   \frac{h_\nu^2 v^2 M^2 \Gamma^{ }_{\!u}}
 {2[(M^2 + 2 \omega^{ } b)^2 + (\omega^{ } \Gamma^{ }_{\! u})^2]}
 \;, \la{Qminus} 
 \ee
 where $h^{ }_\nu$ is a neutrino Yukawa coupling, 
 $v$ is the Higgs expectation value, 
 $\omega \equiv \sqrt{k^2 + M^2}$, 
 $M$ is the mass of a right-handed neutrino, 
 and $b$ is the correction parametrizing the real part
 of \eq\nr{def_Sigma}.
 } 
It is, however, the {\em first} ultrarelativistic
rate that has been computed up to NLO for thermal 
neutrino physics. Even though it will take time
before all other coefficients are known at the same level, we hope 
that the knowledge of one of them helps to motivate such efforts. 

A leptogenesis computation typically comes with two goals: determining
the baryon asymmetry, which is fixed at $T \sim 130$~GeV when sphaleron
processes switch off~\cite{sphaleron}, and determining 
lepton asymmetries, which continue to be produced when $T < 130$~GeV. 
As the rate that we computed only concerns processes that are active
in the Higgs phase (cf.\ \eq\nr{Qminus}), 
it is no surprise that it has little effect 
on the baryon asymmetry: we only observe a variation 
on the per mille level. The influence on lepton asymmetries is a bit
larger, however by considering the benchmark point  $\filledsquare$
from ref.~\cite{degenerate}, chosen because lepton asymmetry
production continues for a long time in this case, 
we found a reduction of lepton
asymmetries by $\sim 1$\%. Hence it appears that for 
practical applications it is sufficient to use the simple 
leading-order expression for $\Gamma^{ }_{\! u}$ from 
\eq\nr{lo_result}. 

Closing on a conceptual note, 
our computation can be interpreted 
as amounting to determining the exponential fall-off of 
a single light-like Wilson line. 
Even though we have verified the independence 
of the result on
the gauge fixing parameter up to NLO, the observable itself is not
manifestly gauge invariant. Another possible starting
point would be 
to consider a Wilson loop
(like in ref.~\cite{prs} but in the Higgs phase), which is gauge 
invariant. In this setup, the information relevant for us
could be extracted by pulling the sides of the Wilson loop 
far from 
each other ($r^{ }_\perp\to \infty$), and interpreting
the coefficient of the 
resulting decay as {\em twice} the width $\Gamma^{ }_{\!u}/2$
that we are interested in.
Our approach has the technical advantage 
that it could be generalized, graph-by-graph, to determining
the complete NLO self-energy of an active neutrino 
beyond the ultrarelativistic regime, 
even if  the practical implementation of this
generalization is challenging, given that a full 4d computation is required. 

%
\section*{Acknowledgements}

This work was partly supported by the Swiss National Science Foundation
(SNF) under grant 200020B-188712.

%
\appendix
\renewcommand{\thesection}{Appendix~\Alph{section}}
\renewcommand{\thesubsection}{\Alph{section}.\arabic{subsection}}
\renewcommand{\theequation}{\Alph{section}.\arabic{equation}}

%
\section{Contributions from NLO diagrams}

In this appendix we record 
the separate contributions to the function $\Sigma(k^{ }_z)$, 
defined in \eq\nr{Sigma}. For momentum dependence, 
specifically $\Sigma'(0)$, it is sufficient to remain at 1-loop (``LO'') 
level (cf.\ \se\ref{ss:wave_fcn}), whereas $\Sigma(0)$ is needed up 
to 2-loop level (``NLO''). 
For completeness we list results 
for a general $R^{ }_\xi$ gauge. The gauge parameter appears
in the masses of the longitudinal gauge bosons and Goldstone modes, 
which are denoted by 
\be
 m'{}^2 \;\equiv\; \xi \, m^2
 \;.\la{xi}
\ee
In terms which are not manifestly
IR finite, we denote by $\mQ^{ }$ a fictitious photon mass,
which is taken to zero at the end of the computation. 

%
\subsection{Master functions}
\la{ss:masters}

Denoting 
$
 q = (\vec{q}^{ }_\perp,q^{ }_z)
$,
$
  d \;\equiv\; 3 - 2\epsilon
$,
$
 \int_\vec{q} \equiv \int \frac{{\rm d}^d\vec{q}}{(2\pi)^d}
$, 
and 
$
 \int_{\vec{q}_\perp} \equiv 
 \int \frac{{\rm d}^{d-1}\vec{q}_\perp^{ }}{(2\pi)^{d-1}}
$, 
the NLO result can be expressed in terms of the two master integrals
\ba
  A(m) 
 & \equiv & 
  \int_\vec{q} \frac{1}{q^2 + m^2}
 \; = \; 
  \int_{\vec{q}_\perp} \frac{1}{2 \epsilon^{ }_{q}}
 \; \stackrel{d=3}{=} \;
   - \frac{m}{4\pi}
 \;,  
 \la{A} \\
 B(m^{ }_1,m^{ }_2) 
 & \equiv & 
  \int_\vec{q} \frac{1}{(q^2 + m_1^2) [(p+q)^2 + m_2^2]}
 \; \stackrel{p^{ }_z = 0}{=} \;
  \int_{\vec{q}_\perp}
  \frac{1}{2 \epsilon^{ }_{q1} \epsilon^{ }_{pq2}
  (\epsilon^{ }_{q1} + \epsilon^{ }_{pq2})} 
 \nn 
  & \stackrel{d=3}{=} &
  \frac{i}{8\pi p^{ }_\perp} 
  \ln\frac{m^{ }_1 + m^{ }_2 - i p^{ }_\perp}
          {m^{ }_1 + m^{ }_2 + i p^{ }_\perp}
 \; = \;
 \frac{1}{4\pi p^{ }_\perp} 
 \arctan\Bigl( \frac{p^{ }_\perp}{m^{ }_1 + m^{ }_2} \Bigr) 
 \;,
 \la{B}
\ea
where 
$
 \epsilon^2_{q1} \equiv q_\perp^2 + m_1^2
$
and 
$
 \epsilon^2_{pq2} \equiv (p_\perp^{ } + q^{ }_\perp)^2 + m_2^2
$.
Sometimes we need the mass derivatives of these functions, defined
as 
\be
 \dot{A}(m) \; \equiv \; \frac{{\rm d} A(m)}{{\rm d} m^2}
 \;, \quad
 \dot{B}(m^{ }_1,m^{ }_2) \; \equiv \; 
 \biggl( 
   \frac{\partial}{\partial m_1^2} + 
   \frac{\partial}{\partial m_2^2}
 \biggr) B(m^{ }_1,m^{ }_2)
 \;. \la{derivatives}
\ee
In addition it is convenient to define the tensor integral 
\be
 B^{ }_{ij} \; \equiv \; 
 \int_\vec{q} \frac{q^{ }_i q^{ }_j}
 {(q^2 + m_1^2)[(p+q)^2 + m_2^2]}
 \; \equiv \;
 \biggl( \delta^{ }_{ij} - \frac{p^{ }_i p^{ }_j}{p^2} \biggr) 
 \, B^{ }_{\T}
 \; + \; 
 \frac{p^{ }_i p^{ }_j}{p^2}
 B^{ }_{\sL}
 \;. \la{Bij} 
\ee
In practice we need $B^{ }_{zz}$ at $p^{ } _z = 0$, 
which is then given by 
$
 B^{ }_{zz} |^{ }_{p^{ }_z = 0} = B^{ }_\T
$.
It is possible to express $B^{ }_\T$ in terms of the integrals
in \eqs\nr{A} and \nr{B}, as
\ba
 B^{ }_\T(m^{ }_1,m^{ }_2) & = & 
 \frac{1}{4(d-1)p_\perp^2}
 \biggl\{ 
    \bigl[ p_\perp^2 - m_1^2 + m_2^2 \bigr] \, A(m^{ }_1) 
  + 
    \bigl[ p_\perp^2 + m_1^2 - m_2^2 \bigr] \, A(m^{ }_2) 
 \nn 
 &  & \quad - \,  
    \bigl[ 
     p_\perp^4 + 2 p_\perp^2(m_1^2 + m_2^2) 
    + (m_1^2 - m_2^2)^2
    \bigr] \, B(m^{ }_1,m^{ }_2)
 \biggr\} 
 \;, \la{BT_rel}
\ea
however it is more compact to display results 
in terms of $B^{ }_\T$, 
thereby avoiding inverse powers of $p_\perp^2$ and $d-1$. 
We note that $B^{ }_\T$ is symmetric in $m^{ }_1\leftrightarrow m^{ }_2$.

For understanding the UV asymptotics of the result, 
we need an expansion of the master integrals in inverse
powers of $p_\perp^2$. Whereas $A$ and $\dot{A}$ are
independent of $p^2_\perp$, for $B$ and $\dot{B}$
from \eqs\nr{B} and \nr{derivatives} these
limiting behaviours read
\ba
  B(m^{ }_1,m^{ }_2)  & \approx &  
  B(0,0) + \frac{A(m^{ }_1) + A(m^{ }_2)}{p_\perp^2}
  + \rmO\Bigl( \frac{1}{p_\perp^4} \Bigr)
  \;, \quad
  B(0,0) \; \stackrel{d=3}{=} \; \frac{1}{8 p^{ }_\perp} 
  \;, \la{B_asymp} \\ 
  \dot{B}(m^{ }_1,m^{ }_2) & \approx &  
  \frac{\dot{A}(m^{ }_1) + \dot{A}(m^{ }_2)}{p_\perp^2}
  + \rmO\Bigl( \frac{1}{p_\perp^4} \Bigr)
  \;. \la{dB_asymp} 
\ea

%
\subsection{$Z^0_{ }$ self-energy}

The $Z^0_{ }$ self-energy diagrams can be depicted as
\begin{eqnarray}
&& 
 \hspace*{-1.0cm}
 \yprocA
 \hspace*{0.50cm}
 \yprocB
 \quad \;, 
 \nn[-10mm]
\end{eqnarray}
where the self-energies are
\begin{eqnarray}
 \selfSmaster
 \hspace*{-0.3cm}
& = & 
 \selfA
 \hspace*{0.00cm}
 \selfB
 \hspace*{0.00cm}
 \selfC
 \hspace*{0.00cm}
 \selfD
 \hspace*{0.00cm}
 \selfE
 \hspace*{0.00cm}
 \selfF
 \hspace*{0.00cm}
\nn
 && 
 \selfG
 \hspace*{0.00cm}
 \selfH
 \hspace*{0.00cm}
 \selfI
 \hspace*{0.00cm}
 \selfJ
 \hspace*{0.00cm}
 \selfK
 \hspace*{0.00cm}
 \selfL
 \quad \;, \la{diagsS}
\\
 \selfTmaster
 \hspace*{-0.3cm}
& = & 
 \selfTA
 \hspace*{0.00cm}
 \selfTC
 \hspace*{0.00cm}
 \selfTE
 \hspace*{0.00cm}
 \selfTF
 \hspace*{0.00cm}
 \selfTI
 \hspace*{0.00cm}
 \selfTJ
 \hspace*{0.00cm}
 \selfTK
 \hspace*{0.00cm}
 \selfTL
 \;. \la{diagsT}
\end{eqnarray}
Here a wiggly line represents a spatial gauge field, 
a dotted line a ghost, 
a dashed line a Higgs field, 
and a solid line a temporal gauge field.
The result reads
\ba
 \Sigma(0) & \supset & \frac{\tilde{g}^4 T^2}{4} \frac{i}{2}
 \int_{\vec{p}_\perp} 
 \biggl\{ \; 
    \frac{\Pi^{\Z}_{\T}}{(p_\perp^2 + \mZ^2)^2}
 \nn 
 & & \quad - \, 
    \frac{\cos^2(\theta-\tilde\theta)\,\Pi^{\Zt}_{ }}{(p_\perp^2 + \mZt^2)^2}
 -  \frac{\sin^2(\theta-\tilde\theta)\,\Pi^{\Qt}_{ }}{(p_\perp^2 + \mQt^2)^2}
 -  \frac{\sin[2(\theta-\tilde\theta)]\,
          \Pi^{\Zt\Qt}_{ }}{2(p_\perp^2 + \mZt^2)(p_\perp^2 + \mQt^2)}
 \; \biggr\}
 \;.
\ea
The transverse spatial self-energy reads 
\ba
 \Pi^{\Z}_{\T} & = & 
  A(\mH^{ }) \, \biggl[ \frac{1}{2} \biggr]
 \nn 
 & + &  
 A(\mZ^{ }) \, \biggl[ \frac{(d-1)\mZ^2}{2\mH^2} \biggr]
 \nn 
 & + &  
 A(\mW^{ }) \, \biggl[ 2(2-d) \cos^4(\theta)
   - \frac{2 (p_\perp^2 + \mW^2) \cos^2(\theta)}{\mZ^2}
                              + \frac{(d-1)\mW^2}{\mH^2} \biggr]
 \nn 
 & + &  
 A(\mW') \, \biggl[ 2\cos^2(\theta)
      \,\frac{p_\perp^2 + \mZ^2 }{\mZ^2} 
            \biggr]
 \nn
 & + &  
 A(\mWt) \, \biggl[ -2 \cos^4(\theta) + \frac{\mW^2}{\mH^2} \biggr]
 \nn
 & + &  
 A(\mZt) \, \biggl[ \frac{\mZ^2 \cos^2(\theta-\tilde{\theta})}{2\mH^2} \biggr]
 \nn
 & + &  
 A(\mQt) \, \biggl[ \frac{\mZ^2 \sin^2(\theta-\tilde{\theta})}{2\mH^2} \biggr]
 \nn
 & + & 
  B(\mH^{ },\mZ^{ }) \, \Bigl[ \mZ^2  \Bigr]
 \nn
 & + & 
  B(\mW^{ },\mW^{ }) \, \biggl[  
   \frac{2 p_\perp^2 (p_\perp^2 + 4 \mW^2) \cos^2(\theta)}{\mZ^2}  \biggr]
 \nn
 & + & 
  B(\mW^{ },\mW')\, \biggl[ 
 -  \frac{2 (p_\perp^2+ \mZ^2) 
            (p_\perp^2 + 2 \mW^2 - \mZ^2)
 \cos^2(\theta)}{\mZ^2}  \biggr]
 \nn 
 & + & 
 B^{ }_\T(\mH^{ },\mZ^{ })\, \bigl[ 1 \bigr]
 \nn 
 & + & 
 B^{ }_\T(\mW^{ },\mW^{ }) \, \biggl[ 
   4(d-2) \cos^4(\theta) + 
   \frac{(p_\perp^2 + 2 \mW^2)^2}{\mZ^4 }
 \biggr]
 \nn 
 & + & 
 B^{ }_\T(\mW^{ },\mW')\, \biggl[ 
 - \frac{2 (p_\perp^2+ \mZ^2) 
           (p_\perp^2 + 2 \mW^2 - \mZ^2)}{\mZ^4} \biggr]
 \nn 
 & + & 
 B^{ }_\T(\mW',\mW')\, \biggl[   
  \frac{p_\perp^4 - \mZ^4}{\mZ^4} \biggr]
 \nn 
 & + & 
 B^{ }_\T(\mWt,\mWt)\, \Bigl[ 4 \cos^4(\theta) \Bigr]
 \;, 
\ea
whereas the temporal parts can be expressed as 
\ba
 \Pi^{\Zt}_{ } \!& = &\! 
  A(\mH^{ }) \, \biggl[ \frac{\cos^2(\theta-\tilde\theta)}{2} \biggr]
 \nn 
 \!& + &\!  
 A(\mZ^{ }) \, \biggl[ \frac{(d-1)\mZ^2\cos^2(\theta-\tilde\theta)}{2\mH^2}
 \biggr]
 \nn 
 \!& + &\!  
 A(\mW^{ }) \, \biggl[ 2(2-d) \cos^2(\theta)\cos^2(\tilde\theta)
   - \frac{2 (p_\perp^2 + \mWt^2) \cos^2(\tilde\theta)}{\mZ^2}
   + \frac{(d-1)\mW^2 \cos^2(\theta-\tilde\theta)}{\mH^2} \biggr]
 \nn 
 \!& + &\!  
 A(\mW') \, \biggl[ 
      \frac{2(p_\perp^2 + \mZt^2)\cos^2(\tilde\theta)}{\mZ^2} 
            \biggr]
 \nn
 \!& + &\!  
 A(\mWt) \, \biggl[ -2 \cos^2(\theta)\cos^2(\tilde\theta)
   + \frac{\mW^2 \cos^2(\theta-\tilde\theta) }{\mH^2} \biggr]
 \nn
 \!& + &\!  
 A(\mZt) \, \biggl[ \frac{\mZ^2 \cos^4(\theta-\tilde{\theta})}{2\mH^2} \biggr]
 \nn
 \!& + &\!  
 A(\mQt) \, \biggl[ \frac{\mZ^2 
 \cos^2(\theta-\tilde\theta) \sin^2(\theta-\tilde{\theta})}{2\mH^2} \biggr]
 \nn
 \!& + &\! 
  B(\mH^{ },\mZt^{ }) \, \Bigl[ \mZ^2 \cos^4(\theta-\tilde\theta) \Bigr]
 \nn
 \!& + &\! 
  B(\mH^{ },\mQt^{ }) \, \Bigl[ 
    \mZ^2 \cos^2(\theta-\tilde\theta) \sin^2(\theta-\tilde\theta) \Bigr]
 \nn
 \!& + &\! 
  B(\mWt^{ },\mW^{ }) \, \biggl[  
   \frac{2 (p_\perp^4 + 2 p_\perp^2(\mW^2 + \mWt^2)
   + (\mW^2 - \mWt^2)^2 ) \cos^2(\tilde\theta)}{\mZ^2}  \biggr]
 \nn
 \!& + &\! 
  B(\mWt^{ },\mW')\, \biggl[ 
 -  \frac{2 (p_\perp^2 + \mZt^2)
            (p_\perp^2 + 2 \mWt^2 - \mZt^2) 
 \cos^2(\tilde\theta)}{\mZ^2}  \biggr]
 \;, \\
 \Pi^{\Zt\Qt}_{ }
 \!& = &\! 
  A(\mH^{ }) \, \biggl[ 
   \frac{\sin(2(\theta-\tilde\theta))}{2} \biggr]
 \nn 
 \!& + &\!  
 A(\mZ^{ }) \, \biggl[ \frac{(d-1)\mZ^2
   \sin(2(\theta-\tilde\theta))}{2\mH^2}
 \biggr]
 \nn 
 \!& + &\!  
 A(\mW^{ }) \, \biggl[ 2(d-2) \cos^2(\theta)
     \sin(2\tilde\theta)
   + \frac{2 (p_\perp^2 + \mWt^2) 
      \sin(2\tilde\theta)}{\mZ^2}
   + \frac{(d-1)\mW^2 
     \sin(2(\theta-\tilde\theta))}{\mH^2} \biggr]
 \nn 
 \!& + &\!  
 A(\mW') \, \biggl[ 
     - \frac{(2 p_\perp^2 + \mZt^2 + \mQt^2)
       \sin(2\tilde\theta)}{\mZ^2} 
            \biggr]
 \nn
 \!& + &\!  
 A(\mWt) \, \biggl[ 2 \cos^2(\theta)\sin(2\tilde\theta)
   + \frac{\mW^2 
   \sin(2(\theta-\tilde\theta)) }{\mH^2} \biggr]
 \nn
 \!& + &\!  
 A(\mZt) \, \biggl[ \frac{\mZ^2 \cos^2(\theta-\tilde{\theta})
 \sin(2(\theta-\tilde\theta))}{2\mH^2} \biggr]
 \nn
 \!& + &\!  
 A(\mQt) \, \biggl[ \frac{\mZ^2 
 \sin^2(\theta-\tilde\theta) \sin(2(\theta-\tilde{\theta}))}{2\mH^2} \biggr]
 \nn
 \!& + &\! 
  B(\mH^{ },\mZt^{ }) \, \Bigl[ \mZ^2 \cos^2(\theta-\tilde\theta)
   \sin(2(\theta - \tilde\theta)) \Bigr]
 \nn
 \!& + &\! 
  B(\mH^{ },\mQt^{ }) \, \Bigl[ 
    \mZ^2 \sin^2(\theta-\tilde\theta) \sin(2(\theta-\tilde\theta)) \Bigr]
 \nn
 \!& + &\! 
  B(\mWt^{ },\mW^{ }) \, \biggl[  
   - \frac{2 (p_\perp^4 + 2 p_\perp^2(\mW^2 + \mWt^2)
   + (\mW^2 - \mWt^2)^2 ) \sin(2\tilde\theta)}{\mZ^2}  \biggr]
 \nn
 \!& + &\! 
  B(\mWt^{ },\mW')\, \biggl[ 
   \frac{2 (p_\perp^4 + 2p_\perp^2 \mWt^2 
    + \mWt^2(\mZt^2 + \mQt^2) - \mQt^2 \mZt^2 ) 
 \sin(2\tilde\theta)}{\mZ^2}  \biggr]
 \;, \\
 \Pi^{\Qt}_{ } 
 \!& = &\! 
  A(\mH^{ }) \, \biggl[ \frac{\sin^2(\theta-\tilde\theta)}{2} \biggr]
 \nn 
 \!& + &\!  
 A(\mZ^{ }) \, \biggl[ \frac{(d-1)\mZ^2\sin^2(\theta-\tilde\theta)}{2\mH^2}
 \biggr]
 \nn 
 \!& + &\!  
 A(\mW^{ }) \, \biggl[ 2(2-d) \cos^2(\theta)\sin^2(\tilde\theta)
   - \frac{2 (p_\perp^2 + \mWt^2) \sin^2(\tilde\theta)}{\mZ^2}
   + \frac{(d-1)\mW^2 \sin^2(\theta-\tilde\theta)}{\mH^2} \biggr]
 \nn 
 \!& + &\!  
 A(\mW') \, \biggl[ 
       \frac{2(p_\perp^2 + \mQt^2)\sin^2(\tilde\theta)}{\mZ^2} 
            \biggr]
 \nn
 \!& + &\!  
 A(\mWt) \, \biggl[ -2 \cos^2(\theta)\sin^2(\tilde\theta)
   + \frac{\mW^2 \sin^2(\theta-\tilde\theta) }{\mH^2} \biggr]
 \nn
 \!& + &\!  
 A(\mZt) \, \biggl[ \frac{\mZ^2 \cos^2(\theta-\tilde{\theta})
 \sin^2(\theta - \tilde\theta)}{2\mH^2} \biggr]
 \nn
 \!& + &\!  
 A(\mQt) \, \biggl[ \frac{\mZ^2 
  \sin^4(\theta-\tilde{\theta})}{2\mH^2} \biggr]
 \nn
 \!& + &\! 
  B(\mH^{ },\mZt^{ }) \, \Bigl[ \mZ^2 \cos^2(\theta-\tilde\theta)
  \sin^2(\theta - \tilde\theta) \Bigr]
 \nn
 \!& + &\! 
  B(\mH^{ },\mQt^{ }) \, \Bigl[ 
    \mZ^2 \sin^4(\theta-\tilde\theta) \Bigr]
 \nn
 \!& + &\! 
  B(\mWt^{ },\mW^{ }) \, \biggl[  
   \frac{2 (p_\perp^4 + 2 p_\perp^2(\mW^2 + \mWt^2)
   + (\mW^2 - \mWt^2)^2 ) \sin^2(\tilde\theta)}{\mZ^2}  \biggr]
 \nn
 \!& + &\! 
  B(\mWt^{ },\mW')\, \biggl[ 
 -  \frac{2 (p_\perp^2 + \mQt^2)
   (p_\perp^2 + 2 \mWt^2 - \mQt^2) 
 \sin^2(\tilde\theta)}{\mZ^2}  \biggr]
 \;.
\ea

%
\subsection{$W^{\pm}_{ }$ self-energy}

The contribution of the $W^\pm_{ }$ self-energy diagrams, {\it viz.}\
\begin{eqnarray}
&& 
 \hspace*{-1.0cm}
 \yprocC
 \hspace*{0.50cm}
 \yprocD
 \quad \;, 
 \nn[-5mm]
\end{eqnarray}
can be written as
\be
 \Sigma(0) \supset \frac{\tilde{g}^4 T^2 \cos^4(\theta)}{2} \frac{i}{2}
 \int_{\vec{p}_\perp} 
 \biggl\{ \;
   \frac{\Pi^{\W}_{\T}}{(p_\perp^2 + \mW^2)^2}
 - 
   \frac{\Pi^{\Wt}_{ }}{(p_\perp^2 + \mWt^2)^2}
 \; \biggr\} 
 \;.
\ee
The transverse spatial self-energy reads 
\ba
 \Pi^{\W}_{\T} & = & 
  A(\mH^{ }) \, \biggl[ \frac{1}{2} \biggr]
 \nn 
 & + &  
 A(\mZ^{ }) \, \biggl[
  \biggl( 2-d 
  -  \frac{p_\perp^2 + \mW^2}{\mZ^2} \biggr) \cos^2(\theta)
  +  \frac{(d-1)\mZ^2}{2\mH^2} \biggr]
 \nn 
 & + &  
 A(\mZ') \, \biggl[
   \frac{(p_\perp^2 + \mW^2)\cos^2(\theta)}{\mZ^2}
   \biggr]
 \nn 
 & + &  
 A(\mW^{ }) \, \biggl[  2 - d 
                         - \frac{p_\perp^2+\mW^2}{\mW^2}
                              + \frac{(d-1)\mW^2}{\mH^2} \biggr]
 \nn 
 & + &  
 A(\mW') \, \biggl[ 
                   \frac{p_\perp^2 + \mW^2}{\mW^2}
             \biggr]
 \nn
 & + &  
 A(\mWt) \, \biggl[  - 1 
                     + \frac{\mW^2}{\mH^2} \biggr]
 \nn
 & + &  
 A(\mZt) \, \biggl[  -\cos^2(\tilde\theta) 
    + \frac{\mZ^2 \cos^2(\theta-\tilde{\theta})}{2\mH^2} \biggr]
 \nn
 & + &  
 A(\mQt) \, \biggl[ -\sin^2(\tilde\theta) 
    +  \frac{\mZ^2 \sin^2(\theta-\tilde{\theta})}{2\mH^2} \biggr]
 \nn
 & + & 
  B(\mH^{ },\mW^{ }) \, \Bigl[  \mW^2  \Bigr]
 \nn
 & + & 
  B(\mZ^{ },\mW^{ }) \, \biggl[ 
   (\mW^2 + \mZ^2 )\, 
     \frac{p_\perp^4 + 2 p_\perp^2 ( \mW^2 + \mZ^2 ) + 
     (\mW^2 - \mZ^2)^2 }{\mZ^4}
  \biggr]
 \nn
 & + & 
  B(\mZ^{ },\mW')\, \biggl[  
   - \frac{(p_\perp^2 + \mW^2)(p_\perp^2 + 2 \mZ^2 - \mW^2)}
 {\mZ^2}  \biggr]
 \nn 
 & + & 
  B(\mZ',\mW^{ })\, \biggl[
   - \frac{(p_\perp^2 + \mW^2)^2\cos^2(\theta)}{\mZ^2}  \biggr]
 \nn 
 & + & 
  B(0,\mW^{ })\, \biggl[
   \frac{(p_\perp^4 + 4 p_\perp^2 \mW^2 - \mW^4)\sin^2(\theta)}{\mW^2}
   \biggr]
 \nn 
 & + & 
  B(0,\mW')\, \biggl[
     \frac{(\mW^4 - p_\perp^4)\sin^2(\theta)}{\mW^2}
  \biggr]
 \nn 
 & + & 
 \biggl[ 
 \frac{  B(\mW^{ },\mQ^{ }) - B(\mW^{ },\mQ') }{\mQ^2}
 + \frac{ A(\mQ') - A(\mQ^{ }) }{ \mQ^2 (p_\perp^2 + \mW^2) }
 \biggr]
 \, \Bigl[ 
  (p_\perp^2 + \mW^2)^2
  \sin^2(\theta)  \Bigr]
 \nn 
 & + & 
 B^{ }_\T(\mH^{ },\mW^{ })\, \bigl[
   1 \bigr]
 \nn 
 & + & 
 B^{ }_\T(\mZ^{ },\mW^{ }) \, \biggl[ 
     4(d-2) \cos^2(\theta) 
   + \frac{p_\perp^4 + 2 p_\perp^2(\mW^2 + \mZ^2) 
  + (\mW^2 + \mZ^2)^2 }{\mZ^4 }
 \biggr]
 \nn 
 & + & 
 B^{ }_\T(\mZ^{ },\mW')\, \biggl[
   - \frac{(p_\perp^2 + \mW^2)(p_\perp^2 + 2 \mZ^2 - \mW^2)}
  {\mZ^4} \biggr]
 \nn 
 & + & 
 B^{ }_\T(\mZ',\mW^{ })\, \biggl[
   - \frac{(p_\perp^2 + \mW^2)^2}{\mZ^4} \biggr]
 \nn 
 & + & 
 B^{ }_\T(\mW',\mZ')\, \biggl[   
    \frac{p_\perp^4 - \mW^4}{\mZ^4} \biggr]
 \nn 
 & + & 
 B^{ }_\T(0,\mW^{ })\, \biggl[
    4(d-2)\sin^2(\theta) 
   + \frac{2 (p_\perp^2 + \mW^2)\sin^2(\theta)}{\mW^2} \biggr]
 \nn 
 & + & 
 B^{ }_\T(0,\mW')\, \biggl[
   - \frac{2 (p_\perp^2 + \mW^2)\sin^2(\theta)}{\mW^2} \biggr]
 \nn 
 & + & 
 B^{ }_\T(\mZt,\mWt)\, \Bigl[
    4 \cos^2(\tilde\theta) \Bigr]
 \nn 
 & + & 
 B^{ }_\T(\mQt,\mWt)\, \Bigl[
    4 \sin^2(\tilde\theta) \Bigr]
 \nn 
 & + & 
 \frac{ B^{ }_\T(\mW',\mQ^{ }) - B^{ }_\T(\mW',\mQ') }{\mQ^2} \, \biggl[    
  \frac{(\mW^4 - p_\perp^4)\sin^2(\theta)}{\mW^2}  \biggr]
 \nn 
 & + & 
 \frac{ B^{ }_\T(\mW^{ },\mQ^{ }) - B^{ }_\T(\mW^{ },\mQ') }{\mQ^2} \, \biggl[
  \frac{(p_\perp^2 + \mW^2)^2  \sin^2(\theta) }{\mW^2} \biggr]
 \;, 
\ea
whereas the temporal part can be expressed as 
\ba
 \Pi^{\Wt}_{ } & = & 
  A(\mH^{ }) \, \biggl[ \frac{1}{2} \biggr]
 \nn 
 & + &  
 A(\mZ^{ }) \, \biggl[
  \biggl( 2-d 
  -  \frac{p_\perp^2 + \mWt^2}{\mZ^2} \biggr) \cos^2(\theta)
  +  \frac{(d-1)\mZ^2}{2\mH^2} \biggr]
 \nn 
 & + &  
 A(\mZ') \, \biggl[
   \frac{(p_\perp^2 + \mWt^2)\cos^2(\theta)}{\mZ^2}
   \biggr]
 \nn 
 & + &  
 A(\mW^{ }) \, \biggl[  2 - d 
             - \frac{(p_\perp^2+\mZt^2)\cos^2(\tilde\theta)}{\mW^2}
             - \frac{(p_\perp^2+\mQt^2)\sin^2(\tilde\theta)}{\mW^2}
                         + \frac{(d-1)\mW^2}{\mH^2} \biggr]
 \nn 
 & + &  
 A(\mW') \, \biggl[ 
                \frac{p_\perp^2 + \mWt^2}{\mW^2}
             \biggr]
 \nn
 & + &  
 A(\mWt) \, \biggl[  - 1 
                     + \frac{\mW^2}{\mH^2} \biggr]
 \nn
 & + &  
 A(\mZt) \, \biggl[  -\cos^2(\tilde\theta) 
    + \frac{\mZ^2 \cos^2(\theta-\tilde{\theta})}{2\mH^2} \biggr]
 \nn
 & + &  
 A(\mQt) \, \biggl[ -\sin^2(\tilde\theta) 
    +  \frac{\mZ^2 \sin^2(\theta-\tilde{\theta})}{2\mH^2} \biggr]
 \nn
 & + & 
  B(\mH^{ },\mWt^{ }) \, \Bigl[  \mW^2  \Bigr]
 \nn
 & + & 
  B(\mZ^{ },\mWt^{ }) \, \biggl[ 
     \frac{(p_\perp^4 + 2 p_\perp^2 ( \mWt^2 + \mZ^2 ) + 
     (\mWt^2 - \mZ^2)^2 )\cos^2(\theta)}{\mZ^2}
  \biggr]
 \nn
 & + & 
  B(\mZ',\mWt^{ })\, \biggl[
   - \frac{(p_\perp^2 + \mWt^2)^2\cos^2(\theta)}{\mZ^2}  \biggr]
 \nn 
 & + & 
  B(\mZt^{ },\mW^{ }) \, \biggl[ 
     \frac{(p_\perp^4 + 2 p_\perp^2 ( \mW^2 + \mZt^2 ) + 
     (\mW^2 - \mZt^2)^2)\cos^2(\tilde\theta)}{\mW^2}
  \biggr]
 \nn
 & + & 
  B(\mQt^{ },\mW^{ }) \, \biggl[ 
     \frac{(p_\perp^4 + 2 p_\perp^2 ( \mW^2 + \mQt^2 ) + 
     (\mW^2 - \mQt^2)^2)\sin^2(\tilde\theta)}{\mW^2}
  \biggr]
 \nn
 & + & 
  B(\mZt^{ },\mW')\, \biggl[  
   - \frac{(p_\perp^2 + \mWt^2)
   (p_\perp^2 + 2 \mZt^2 - \mWt^2)  
  \cos^2(\tilde\theta)}
 {\mW^2}  \biggr]
 \nn 
 & + & 
  B(\mQt^{ },\mW')\, \biggl[  
   - \frac{(p_\perp^2 + \mWt^2)
   (p_\perp^2 + 2 \mQt^2 - \mWt^2) 
   \sin^2(\tilde\theta)}
 {\mW^2}  \biggr]
 \nn 
 & + & 
 \biggl[ 
 \frac{  B(\mWt^{ },\mQ^{ }) - B(\mWt^{ },\mQ') }{\mQ^2}
 + \frac{ A(\mQ') - A(\mQ^{ }) }{ \mQ^2 (p_\perp^2 + \mWt^2) }
 \biggr]
 \, \Bigl[ 
  (p_\perp^2 + \mWt^2)^2
  \sin^2(\theta)  \Bigr]
 \nn 
 & + & 
  B(0,\mWt^{ })\, \Bigl[
   2 (p_\perp^2 - \mWt^2) \sin^2(\theta)
   \Bigr]
 \;.  
\ea

%
\subsection{Triple gauge vertex}

The contribution of triple gauge vertex diagrams, {\it viz.}\
\begin{eqnarray}
&& 
 \hspace*{-1.0cm}
 \yprocEa
 \hspace*{0.50cm}
 \yprocEb
 \hspace*{0.50cm}
 \yprocEc
 \hspace*{0.50cm}
 \yprocFa
 \hspace*{0.50cm}
 \yprocFb
 \hspace*{0.50cm}
 \yprocFc
 \nn
&& 
 \hspace*{-1.0cm}
 \yprocGa
 \hspace*{0.50cm}
 \yprocHb
 \hspace*{0.50cm}
 \yprocGb
 \hspace*{0.50cm}
 \yprocHa
 \hspace*{0.50cm}
 \yprocGc
 \hspace*{0.50cm}
 \yprocHc
 \quad \;, 
 \nn[-10mm]
\end{eqnarray}
can be written as
\ba
 \Sigma(0)  & \supset &    
 \frac{\tilde{g}^4 T^2 \cos^3(\theta)}{2} \frac{i}{2}
 \int_{\vec{p}_\perp} 
 \biggl\{ \frac{\Upsilon^\Z_{ }} {p_\perp^2 + \mZ^2}
 + \frac{\Upsilon^\W_{ }} {p_\perp^2 + \mW^2}
 \nn 
 & & \quad 
 - \, \frac{\Upsilon^\Zt_{ }} {p_\perp^2 + \mZt^2}
 - \frac{\Upsilon^\Qt_{ }} {p_\perp^2 + \mQt^2}
 - \frac{\Upsilon^\Wt_{ }} {p_\perp^2 + \mWt^2}
 \biggr\} 
 \;,  
\ea
where
\ba
 \Upsilon^\Z_{ } & = & \cos(\theta) \,
 \Bigl[  \theta^{ }_{zz}(\mW^{ },\mW^{ }) 
        + 2 B(\mWt^{} ,\mWt^{ })
 \Bigr]
 \;,  \\ 
 \Upsilon^\W_{ } & = & 
    2 \cos(\theta) \,
  \Bigl[ \cos^2(\theta) \theta^{ }_{zz}(\mZ^{ },\mW^{ })
  + 2 \cos^2(\tilde\theta) B(\mZt^{ },\mWt^{ }) 
  \Bigr]
 \nn 
 & + & 
   2 \cos(\theta) \, 
  \Big[ \sin^2(\theta) \theta^{ }_{zz}(\mQ^{ },\mW^{ })
  + 2 \sin^2(\tilde\theta) B(\mQt^{ },\mWt^{ })
  \Bigr]
 \;, \\ 
 \Upsilon^\Zt_{ } & = &
 2 \cos(\tilde\theta) \cos(\theta - \tilde \theta)
 \, \theta^{ }_{z0}(\mW^{ },\mWt^{ })
 \;, \\ 
 \Upsilon^\Qt_{ } & = &
 - 
 2 \sin(\tilde\theta) \sin(\theta - \tilde \theta)
 \, \theta^{ }_{z0}(\mW^{ },\mWt^{ })
 \;,  \\ 
 \Upsilon^\Wt_{ } & = & 
 2 \cos(\theta) \Bigl[ 
  \cos^2(\theta) \theta^{ }_{z0}(\mZ^{ },\mWt^{ })
 + \cos^2(\tilde\theta)  \theta^{ }_{z0}(\mW^{ },\mZt^{ })
  \Bigr]
 \nn  
 & + & 
 2 \cos(\theta) \Bigl[ 
  \sin^2(\theta)  \theta^{ }_{z0}(\mQ^{ },\mWt^{ })
+ \sin^2(\tilde\theta)  \theta^{ }_{z0}(\mW^{ },\mQt^{ })
 \Bigr]
 \;.
\ea
Here we have denoted 
\ba
 \theta_{zz}^{ }(m^{ }_1,m^{ }_2) & \equiv &
 \int_\vec{q} 
 \mathbbm{P}
 \biggl\{ 
 \frac{\gamma^{ }_{3ij}(p,q)}{q^{ }_z}
 \, \Delta^{ }_{3i}(q,m^{ }_1)
    \Delta^{ }_{3j}(p+q,m^{ }_2)  
 \biggr\}^{ }_{p^{ }_z = 0}
 \\
 & = &   
   \frac{A(m^{ }_1)- A(m_1')}{m_1^2} 
 + 
   \frac{A(m^{ }_2)- A(m_2')}{m_2^2} 
  +    
  \frac{ p_\perp^2
  [B^{ }_\T(m_1^{ },m_2^{ }) - B^{ }_\T(m_1',m_2') ]}{m_1^2 m_2^2} 
 \nn
 & + &   
  \frac{p_\perp^2 + m_1^2}{m_2^2} 
  \biggl[ B(m^{ }_1,m_2') - B(m^{ }_1,m^{ }_2) 
   + \frac{
  B^{ }_\T(m^{ }_1,m_2') - B^{ }_\T(m^{ }_1,m^{ }_2) }{m_1^2}
  \biggr] 
 \nn
 & + &   
  \frac{p_\perp^2 + m_2^2}{m_1^2} 
  \biggl[ B(m_1',m^{ }_2) - B(m^{ }_1,m^{ }_2) 
 + 
  \frac{
   B^{ }_\T(m_1',m^{ }_2) - B^{ }_\T(m^{ }_1,m^{ }_2) }{m_2^2}
  \biggr]
 \;, \nn \\ 
 \theta_{z0}^{ }(m^{ }_1,m^{ }_2) & \equiv &
 \int_\vec{q} 
 \mathbbm{P}
 \biggl\{ 
 \frac{\tilde\gamma^{ }_{3ij}(p,q)}{q^{ }_z}
 \, \Delta^{ }_{ij}(q,m^{ }_1)
    \Delta^{ }_{00}(p+q,m^{ }_2)  
 \biggr\}^{ }_{p^{ }_z = 0}
 \\
 & = &   
   B(m^{ }_1,m^{ }_2)
   + \frac{A(m^{ }_1)- A(m_1')}{m_1^2} 
 \nn 
  & + & 
  \frac{p_\perp^2 + m_2^2}{m_1^2} 
  \biggl[ B(m_1',m^{ }_2) - B(m^{ }_1,m^{ }_2) 
  \biggr]
 \;,
\ea
where 
\ba 
 \gamma^{ }_{3ij}(p,q) & \equiv & \delta^{ }_{3i}(q^{ }_j - p^{ }_j) 
      + \delta^{ }_{3j}(q^{ }_i + 2 p^{ }_i) 
      - \delta^{ }_{ij}(2 q^{ }_z + p^{ }_z)
 \;, \\[2mm]
 \tilde\gamma^{ }_{3ij}(p,q) & \equiv & 
       \delta^{ }_{3j}(q^{ }_i + 2 p^{ }_i) 
 \;, \\
 \Delta^{ }_{00}(p,m) 
  & \equiv & 
 \frac{1}{p^2 + m^2}
 \;, \la{Delta0} \\
 \Delta^{ }_{ij}(p,m) 
 & \equiv & 
 \frac{\delta^{ }_{ij}}{p^2 + m^2} + 
 \frac{p^{ }_i p^{ }_j}{m^2}
 \biggl(
   \frac{1}{p^2 + m^2} - \frac{1}{p^2 + m'{}^2} 
 \biggr)
 \;. \la{Delta} 
\ea
We note that $\theta_{zz}^{ }$ is symmetric in 
$m^{ }_1\leftrightarrow m^{ }_2$.

%
\subsection{Crossed fermion self-energy}

For the ``crossed'' NLO self-energy diagrams, {\it viz.} 
\begin{eqnarray}
&& 
 \hspace*{-1.0cm}
 \yprocI
 \hspace*{0.50cm}
 \yprocId
 \hspace*{0.50cm}
 \yprocIf
 \hspace*{0.50cm}
 \yprocJ
 \hspace*{0.50cm}
 \yprocJd
 \hspace*{0.50cm}
 \yprocJf
 \nn
&& 
 \hspace*{-1.0cm}
 \yprocK
 \hspace*{0.50cm}
 \yprocKd
 \hspace*{0.50cm}
 \yprocKf
 \hspace*{0.50cm}
 \yprocL
 \hspace*{0.50cm}
 \yprocLd
 \hspace*{0.50cm}
 \yprocLf
 \nn[-10mm]
\end{eqnarray}
the result can be written as 
\ba
 \Sigma(0) & \supset & 
 \frac{\tilde{g}^4 T^2}{16}
 \Bigl\{ 
   \phi^{ }_{zz}(\mZ^{ },\mZ^{ })
 \nn 
 & & \quad - \, 
    2\cos^2(\theta - \tilde\theta) \phi^{ }_{0z}(\mZt^{ },\mZ^{ })
  - 2\sin^2(\theta - \tilde\theta) \phi^{ }_{0z}(\mQt^{ },\mZ^{ })
 \nn 
 & & \quad + \, 
    \cos^4(\theta - \tilde\theta) \phi^{ }_{00}(\mZt^{ },\mZt^{ })
  + \sin^4(\theta - \tilde\theta) \phi^{ }_{00}(\mQt^{ },\mQt^{ })
 \nn 
 & & \quad + \, 
  2 \cos^2(\theta - \tilde\theta) 
    \sin^2(\theta - \tilde\theta) \phi^{ }_{00}(\mZt^{ },\mQt^{ })
 \nn 
  & + &
  4 \cos^2(\theta) \Bigl[ 
     \cos(2\theta)
     \Bigl( 
               \phi^{ }_{z0}(\mZ^{ },\mWt^{ })
            -  \phi^{ }_{zz}(\mZ^{ },\mW^{ })
     \Bigr)
 \nn 
 & & \quad + \, 
     \cos(\theta-\tilde\theta)\cos(\theta+\tilde\theta)
     \Bigl( 
          \phi^{ }_{0z}(\mZt^{ },\mW^{ })
       -  \phi^{ }_{00}(\mZt^{ },\mWt^{ })
     \Bigr)
 \nn 
 & & \quad + \, 
    \sin(\theta-\tilde\theta)\sin(\theta+\tilde\theta)
    \Bigl( 
         \phi^{ }_{00}(\mQt^{ },\mWt^{ })
      -  \phi^{ }_{0z}(\mQt^{ },\mW^{ })
    \Bigr)
 \Bigr]
 \Bigr\} 
 \;, \la{abelian1_master}
\ea
where 
\ba
 \phi^{ }_{ij}(m^{ }_1,m^{ }_2)
 & \equiv & 
 \int_\vec{p,q}
 \frac{\Delta^{ }_{ii}(p,m^{ }_1)\Delta^{ }_{jj}(q,m^{ }_2)}
 {(p^{ }_z - i 0^+)(p^{ }_z + q^{ }_z - i 0^+)(q^{ }_z - i 0^+)}
 \;, 
 \la{abelian1} 
\ea
and the propagators are from \eqs\nr{Delta0} and \nr{Delta}. 
The integrals over $p^{ }_z,q^{ }_z$ can be carried out by contour
integration, most simply by closing in the lower half-plane so that 
the denominators in \eq\nr{abelian1} have no pole, or alternatively 
by inserting \eq\nr{master}. 
For the part $\propto p_z^2$ from \eq\nr{Delta}, this yields contributions
which are directly identified with the master functions $A$, $B$ from
\se\ref{ss:masters}. The other parts require some more
work, either by writing $1/p_z^2 = \partial^{ }_{p_z^{ }}(-1/p_z^{ })$
and carrying out a partial integration, or by resorting to contour
integration. In this way we obtain
($
 \epsilon^{ }_{pi} \equiv \sqrt{p_\perp^2 + m_i^2}\;
$)
\ba
 && \hspace*{-1.0cm}
 \phi^{ }_{00}(m^{ }_1,m^{ }_2)
 \; = \; 
 -\frac{i}{4} 
 \int_\vec{p_\perp,q_\perp}
 \frac{1}{\epsilon^2_{p1} \epsilon^2_{q2}
 (\epsilon^{ }_{p1} + \epsilon^{ }_{q2})}
 = 
 {i} \int_\vec{p_\perp}
 \biggl\{ 
   \frac{ \dot{A}(m^{ }_2) }{p_\perp^2 + m_1^2} 
 + 
   \frac{ \dot{A}(m^{ }_1) }{p_\perp^2 + m_2^2} 
 - 
   \dot{B}(m^{ }_1,m^{ }_2)
 \biggr\} 
 \;, \nn \la{phi00} 
\ea 
where in the last step we substituted integration 
variables and re-expressed the result in terms
of the master integrals from \se\ref{ss:masters}, evaluated 
with $p^{ }_z = 0$.
The other integrals read 
\ba
 \phi^{ }_{0z}(m^{ }_1,m^{ }_2)
 & = & 
 \phi^{ }_{z0}(m^{ }_2,m^{ }_1)
 \nn 
 & = & 
 \phi^{ }_{00}(m^{ }_1,m^{ }_2)
 \nn 
 & + &  
 \frac{i}{2} \int^{ }_{\vec{p}_\perp}
 \biggl\{ \;
  \frac{1}{m_2^2} 
   \biggl[ 
   \frac{A(m^{ }_2) - A(m_2')}{p_\perp^2 + m_1^2}
   + B(m_1^{ },m_2') - B(m_1^{ },m_2^{ })
  \biggr]
 \;
 \biggr\}
 \;, \hspace*{5mm} \\[3mm]  
 \phi^{ }_{zz}(m^{ }_1,m^{ }_2)
 & = & 
 \phi^{ }_{0z}(m^{ }_1,m^{ }_2)
 \nn 
 & + &  
 \frac{i}{2} \int^{ }_{\vec{p}_\perp}
 \biggl\{ 
  \frac{1}{m_1^2} 
  \biggl[ 
   \frac{A(m^{ }_1) - A(m_1')}{p_\perp^2 + m_2^2}
   + B(m_1',m^{ }_2) - B(m_1^{ },m_2^{ })
  \biggr]
 \nn 
 & - & 
 \frac{
  B_\T^{ }(m^{ }_1,m^{ }_2)
  - B_\T^{ }(m_1',m^{ }_2)
  - B_\T^{ }(m^{ }_1,m_2')
  + B_\T^{ }(m_1',m_2')
 }{m_1^2 m_2^2}
 \;
 \biggr\} 
 \;. \la{phizz}
\ea

%
\subsection{Uncrossed fermion self-energy}

The result for the ``uncrossed'' NLO self-energy diagrams, {\it viz.} 
\begin{eqnarray*}
&& 
 \hspace*{-1.0cm}
 \yprocM
 \hspace*{0.50cm}
 \yprocMb
 \hspace*{0.50cm}
 \yprocMe
 \hspace*{0.50cm}
 \yprocMg
 \hspace*{0.50cm}
 \yprocO
 \hspace*{0.50cm}
 \yprocOb
 \hspace*{0.50cm}
 \yprocOe
 \hspace*{0.50cm}
 \yprocOg
 \nn
&& 
 \hspace*{-1.0cm}
 \yprocN
 \hspace*{0.50cm}
 \yprocNb
 \hspace*{0.50cm}
 \yprocNe
 \hspace*{0.50cm}
 \yprocNg
 \hspace*{0.50cm}
 \yprocP
 \hspace*{0.50cm}
 \yprocPb
 \hspace*{0.50cm}
 \yprocPe
 \hspace*{0.50cm}
 \yprocPg
 \nn[-10mm]
\end{eqnarray*}
can be written as 
\ba
 \Sigma(0) & \supset & 
 \frac{\tilde{g}^4 T^2}{16}
 \Bigl\{ 
   \chi^{ }_{zz}(\mZ^{ },\mZ^{ })
     - \cos^2(\theta - \tilde\theta) \chi^{ }_{0z}(\mZt^{ },\mZ^{ })
     - \sin^2(\theta - \tilde\theta) \chi^{ }_{0z}(\mQt^{ },\mZ^{ })
  \nn 
  & & \quad - \,
       \cos^2(\theta - \tilde\theta) \chi^{ }_{z0}(\mZ^{ },\mZt^{ })
     - \sin^2(\theta - \tilde\theta) \chi^{ }_{z0}(\mZ^{ },\mQt^{ })
  \nn 
  & & \quad + \, 
       \cos^4(\theta - \tilde\theta) \chi^{ }_{00}(\mZt^{ },\mZt^{ })
     + \sin^4(\theta - \tilde\theta) \chi^{ }_{00}(\mQt^{ },\mQt^{ })
  \nn 
  & & \quad + \, 
       \cos^2(\theta - \tilde\theta) 
       \sin^2(\theta - \tilde\theta)
       \, \Bigl[ 
         \chi^{ }_{00}(\mZt^{ },\mQt^{ })
         + \chi^{ }_{00}(\mQt^{ },\mZt^{ })
       \Bigr]
 \nn 
  & + &
  2 \cos^2(\theta)\, 
   \Bigl[
      \chi^{ }_{zz}(\mZ^{ },\mW^{ })
     - \cos^2(\theta - \tilde\theta) \chi^{ }_{0z}(\mZt^{ },\mW^{ })
     - \sin^2(\theta - \tilde\theta) \chi^{ }_{0z}(\mQt^{ },\mW^{ })
  \nn 
  & & \quad - \,
       \chi^{ }_{z0}(\mZ^{ },\mWt^{ })
     + \cos^2(\theta - \tilde\theta) \chi^{ }_{00}(\mZt^{ },\mWt^{ })
     + \sin^2(\theta - \tilde\theta) \chi^{ }_{00}(\mQt^{ },\mWt^{ })
   \Bigr]
 \nn 
  & + & 
  4 \cos^4(\theta)\,
   \Bigl[ 
       \chi^{ }_{zz}(\mW^{ },\mW^{ })
      - \chi^{ }_{0z}(\mWt^{ },\mW^{ })
      -  \chi^{ }_{z0}(\mW^{ },\mWt^{ })
      +  \chi^{ }_{00}(\mWt^{ },\mWt^{ })
   \Bigr]
 \nn 
  & + &  
  2 \cos^2(\theta)
   \Bigl[ 
    \cos^2(2\theta)\, \chi^{ }_{zz}(\mW^{ },\mZ^{ })
    - \cos^2(2\theta)\, \chi^{ }_{0z}(\mWt^{ },\mZ^{ })
  \nn 
  & & \quad - \,
      \cos^2(\theta+\tilde\theta) \chi^{ }_{z0}(\mW^{ },\mZt^{ })
    +  \cos^2(\theta+\tilde\theta) \chi^{ }_{00}(\mWt^{ },\mZt^{ })
   \Bigr]
 \nn 
  & + &
  2 \cos^2(\theta)
   \Bigl[ 
    \sin^2(2\theta)\, \chi^{ }_{zz}(\mW^{ },\mQ^{ })
    - \sin^2(2\theta)\, \chi^{ }_{0z}(\mWt^{ },\mQ^{ })
  \nn 
  & & \quad - \,
      \sin^2(\theta+\tilde\theta) \chi^{ }_{z0}(\mW^{ },\mQt^{ })
     + \sin^2(\theta+\tilde\theta) \chi^{ }_{00}(\mWt^{ },\mQt^{ })
   \Bigr]
 \Bigr\} 
 \;, \la{abelian2_master}
\ea
where, making use of  
$\Delta^{ }_{ij}$ defined according to 
\eqs\nr{Delta0} and \nr{Delta},  
\ba
 \chi^{ }_{ij}(m^{ }_1,m^{ }_2)
 & \equiv & 
 \int_\vec{p,q}
 \frac{\Delta^{ }_{ii}(p,m^{ }_1)\Delta^{ }_{jj}(q,m^{ }_2)}
 {(p^{ }_z - i 0^+)^2(p^{ }_z + q^{ }_z - i 0^+)}
 \;.
 \la{abelian2}  
\ea

For dealing with the double pole in \eq\nr{abelian2}, it is convenient
to write 
\ba
 && \hspace*{-2.5cm}
 \frac{1}{(p^{ }_z - i 0^+)^2(p^{ }_z + q^{ }_z - i 0^+)}
 \; = \; 
 - 
 \frac{1}
 {(p^{ }_z - i 0^+)(p^{ }_z + q^{ }_z - i 0^+)(q^{ }_z - i 0^+)}
 \nn 
 & + & 
 \frac{1}
 {(p^{ }_z - i 0^+)(p^{ }_z + q^{ }_z - i 0^+)} 
 \underbrace{
 \biggl( 
   \frac{1}{p^{ }_z - i 0^+}  
 + 
   \frac{1}{q^{ }_z - i 0^+}  
 \biggr) 
 }_
 {\frac{p^{ }_z + q^{ }_z - i 0^+}{(p^{ }_z - i 0^+)(q^{ }_z - i 0^+)}}
 \;. 
\ea
Therefore 
\be
 \chi^{ }_{ij}(m^{ }_1,m^{ }_2) = - \phi^{ }_{ij}(m^{ }_1,m^{ }_2)
 + \delta \chi^{ }_{ij}(m^{ }_1,m^{ }_2)
 \;, \la{chi_phi}
\ee
where $\phi^{ }_{ij}$ is from \eqs\nr{phi00}--\nr{phizz} and 
\ba
 \delta \chi^{ }_{ij}(m^{ }_1,m^{ }_2) & = & 
 \int_\vec{p} \frac{\Delta^{ }_{ii}(p,m^{ }_1)}{(p^{ }_z - i 0^+)^2}
 \int_\vec{q} \frac{\Delta^{ }_{jj}(q,m^{ }_2)}{q^{ }_z - i 0^+}
 \;. 
 \la{chi_ij}
\ea
These integrals can be carried out by contour integration,
or by making use of \eq\nr{master}, noting that $\Delta(q,m^{ }_2)$ is odd
in $q^{ }_z$ so that only the imaginary part contributes, and removing
$1/p_z^2$ from the other term through partial integration, 
like around \eq\nr{phi00}. This yields
\ba 
 \delta \chi^{ }_{00}(m^{ }_1,m^{ }_2)
 & = & 
 \delta \chi^{ }_{0z}(m^{ }_1,m^{ }_2)
 \; = \; 
 \frac{i}{2} 
 \int_\vec{p_\perp} \frac{1}{p_\perp^2 + m_2^2}
 \, 
 \bigl[
 \, 
   2\, \dot{A}(m^{ }_1)
 \, 
 \bigr] 
 \;, \la{delta_chi_00} \\ 
 \delta \chi^{ }_{z0}(m^{ }_1,m^{ }_2)
 & = & 
 \delta \chi^{ }_{zz}(m^{ }_1,m^{ }_2)
 \nn 
 & = & 
 \delta \chi^{ }_{00}(m^{ }_1,m^{ }_2)
 \; + \; 
 \frac{i}{2} 
 \int_\vec{p_\perp} \frac{1}{p_\perp^2 + m_2^2}
 \, 
 \biggl[
 \, 
   \frac{A(m_1^{ }) - A(m_1')}{m_1^2} 
 \, 
 \biggr] 
 \;. \la{delta_chi_zz}
\ea
The contributions from 
$\delta\chi$ are closely related to those in \se\ref{ss:wave_fcn}.

%
\subsection{Wave function normalization}
\la{ss:wave_fcn}

The final contribution originates 
from the second term in \eq\nr{I}, with the various channels of 
\fig\ref{fig:lo} contributing with coefficients like in \eq\nr{lo_result}.
The tree-level part of the self-energy, $k^{ }_z$, gets corrected by this
term (cf.\ \eq\nr{Sigma}), and the correction needs to be factored out, 
in order to determine the location of the pole of the corresponding 
propagator. In other words, we write the combination appearing
in \eq\nr{Sigma} as 
\ba
 k^{ }_z + \Sigma(k^{ }_z)
 & = & k^{ }_z \, \bigl[ 1 + \Sigma'(0) \bigr] + 
 \Sigma(0) + \rmO(k_z^2) 
 \nn 
 & = & 
 \bigl[ 1 + \Sigma'(0) \bigr] \Bigl[ k^{ }_z +
   \frac{\Sigma(0)}{1 + \Sigma'(0)} \Bigr]
 + \rmO(k_z^2) 
 \;. 
\ea
This implies that, up to NNLO corrections,  
the physical width is 
\be
 \frac{\Gamma^{ }_u}{2}
 \approx
 \im\biggl[ \frac{\Sigma(0)}{1 + \Sigma'(0)} \biggr]
 = 
 \im \Sigma^{ }_\rmii{LO}(0) + 
 \bigl\{ \, 
 \im \Sigma^{ }_\rmii{NLO}(0) - 
 \Sigma'_\rmii{LO}(0) \, \im \Sigma^{ }_\rmii{LO}(0)
 \, \bigr\} 
 + 
 \rmO\biggl( \frac{ \tilde{g}^6 T^3}{m_i^2} \biggr)
 \;. 
\ee
The last term shown reads
\ba
 - i\, 
 \Sigma'_\rmii{LO}(0) \, \im \Sigma^{ }_\rmii{LO}(0)
 &  = & -\frac{\tilde{g}^4 T^2}{16} \frac{i}{2} 
 \int_\vec{p_\perp}
 \biggl\{ 
   \frac{1}{p_\perp^2 + \mZ^2} 
  -  
   \frac{\cos^2(\theta - \tilde{\theta})}{p_\perp^2 + \mZt^2}
  -  
   \frac{\sin^2(\theta - \tilde{\theta})}{p_\perp^2 + \mQt^2}
 \nn 
 & & \hspace*{1cm} + \,   
  2\cos^2(\theta)
   \biggl[ \frac{1}{p_\perp^2 + \mW^2}
        -  \frac{1}{p_\perp^2 + \mWt^2}
   \biggr] \biggr\} 
 \nn 
 & \times & 
 \biggl\{ 
  \frac{A(\mZ^{ }) - A(\mZ')}{\mZ^2} 
 + 
  2 \cos^2(\theta) 
  \, \frac{A(\mW^{ }) - A(\mW')}{\mW^2} 
 \nn 
 & & + \, 
  2 \bigl[
       \dot{A}(\mZ^{ })
   - \cos^2(\theta-\tilde{\theta})
       \dot{A}(\mZt^{ })
   - \sin^2(\theta-\tilde{\theta})
       \dot{A}(\mQt^{ })
      \bigr]
 \nn 
 & & + \, 
 4 \cos^2(\theta)
  \bigl[ 
     \dot{A}(\mW^{ })
  -  \dot{A}(\mWt^{ })
  \bigr]
 \biggr\} 
 \;. \hspace*{10mm}
\ea

%

\end{document}